\documentclass[twocolumn,showpacs,preprintnumbers,amsmath,amssymb,superscriptaddress,prb,longbibliography]{revtex4-1}

\usepackage{graphicx}
\usepackage{dcolumn}
\usepackage{bm}
\usepackage{graphics}
\usepackage{subfigure}
\usepackage{graphicx}
\usepackage{epsfig}
\usepackage{epstopdf}
\usepackage{xcolor}
\usepackage{multirow}
\usepackage{mathtools}

\allowdisplaybreaks

\begin{document}

\title{Electromagnetic response of composite Dirac fermions in the half-filled Landau level}

\author{Johannes Hofmann}
\email{johannes.hofmann@physics.gu.se}
\affiliation{Department of Physics, Gothenburg University, 41296 Gothenburg, Sweden}

\date{\today}

\begin{abstract}
An effective field theory of composite Dirac fermions was proposed by Son [Phys.~Rev.~X~{\bf 5}, 031027 (2015)] as a theory of the half-filled Landau level with explicit particle-hole symmetry. We compute the electromagnetic response of this Son-Dirac theory on the level of the random phase approximation (RPA), where we pay particular attention to the effect of an additional composite-fermion dipole term that is needed to restore Galilean invariance. We find that once this dipole correction is taken into account, spurious interband transitions and collective modes that are present in the response of the unmodified theory either cancel or are strongly suppressed. We demonstrate that this gives rise to a consistent theory of the half-filled Landau level valid at all frequencies, at least to leading order in the momentum. In addition, the dipole contribution modifies the Fermi-liquid response at small frequency and momentum, which is a prediction of the Son-Dirac theory within the RPA that distinguishes it from a separate description of the half-filled Landau level by Halperin, Lee, and Read within the RPA.
\end{abstract}

\maketitle

\section{Introduction}

The fractional quantum Hall effect (FQHE) in the lowest Landau level (LLL) is a prototypical example of a strong-interaction phenomenon, where, due to the quenching of the kinetic electron energy in a magnetic field, only a single (interaction) scale remains~\cite{jain07}. Despite the absence of a small parameter, significant progress has been made by describing FQH states in terms of composite fermions, which are quasiparticles formed of electrons and an even number of vortices~\cite{jain89}. The field-theoretical description of composite fermions is based on a Chern-Simons theory that is obtained from the Hamiltonian of interacting electrons in a magnetic field by a formally exact singular gauge transformation, which attaches a number of flux quanta to each electron~\cite{lopez98,giuliani05}. The advantage of this formulation is that  standard many-body approximations --- such as a mean-field approximation for the ground state and a random phase approximation (RPA) for the fluctuations~\cite{simon94a,simon94b,simon98} --- provide an accurate description of the FQHE. In particular, in the special case of the half-filled Landau level, mean-field theory predicts that the Aharonov-Bohm flux attached to each electron precisely cancels the external magnetic field~\cite{tong16}, and the composite fermions form a Fermi liquid, a field-theoretical analysis of which including RPA excitations was first given by Halperin, Lee, and Read (HLR)~\cite{hlr93}. There is considerable experimental evidence for the existence of a compressible state of this form~\cite{willett90,kang93,goldman94}. 

In its original formulation, however, HLR theory includes all electron Landau levels, and must be modified in the LLL limit to account for an effective-mass renormalization (which sets the interaction scale) and to ensure Galilean invariance~\cite{simon93,simon96,simon98}.  A drawback of this modified HLR description is that a particle-hole symmetry --- which is an exact symmetry in the LLL that links the response at filling fractions $\nu$ and $1-\nu$ and constrains the properties of the half-filled Landau level --- is not apparent~\cite{girvin84,kivelson97}. This could mean that calculations within HLR theory have to be carefully revisited to establish consistency with particle-hole symmetry~\cite{wang17,kumar19}, but it could also point to a breakdown of this framework~\cite{levin17,nguyen18,son18}. The latter point was addressed by Son, who proposed an alternative effective Chern-Simons field theory of the half-filled Landau level in terms of composite Dirac fermions~\cite{son15}. The main difference between the Son-Dirac theory and HLR theory lies in the nature of low-energy excitations, with the Dirac composite fermions having an additional Berry phase~\cite{geraedts16,geraedts18,wang19}. Possible discrepancies between the HLR and Son-Dirac theories are of significant current interest, especially since recent experiments have been able to probe the FQHE while varying the electron density independently of a large magnetic field and thus probe the effects of particle-hole symmetry~\cite{pan20}.

Even though the Son-Dirac theory is formulated in terms of composite Dirac fermions, what is not expected in the excitation spectrum is a Dirac cone, especially high-energy features associated with transitions between a Dirac valence and conduction band --- after all, the theory is an effective approximation to the exact theory of nonrelativistic composite fermions, where such a feature is absent~\cite{balram16,halperin20}. The response of the Son-Dirac theory should therefore not resemble the response of typical Dirac materials, such as graphene. However, what ought to be true in principle is not always obvious in direct calculations. In this paper, therefore, we compute the density and current response function of the Son-Dirac theory using the RPA, and we demonstrate how this gives rise to a consistent theory of the half-filled Landau level, at least at long wavelengths. This is not directly apparent since the RPA links the response of the interacting system to the response of noninteracting two-dimensional (2D) Dirac fermions, which naturally includes interband excitations~\cite{hwang07,wunsch06} and collective modes that violate Kohn's theorem~\cite{throckmorton18,hofmann19} (for graphene, for example, these are well-established experimental features~\cite{mak08,ju11,grigorenko12}). Indeed, as we show in this paper,  such spurious contributions are suppressed only if we consider a Galilean-invariant modification of the Son-Dirac theory that includes an additional dipole term for the composite Dirac fermion. The remaining response at low energy and small momentum is then consistent with the predictions of HLR theory, but there are differences in the excitation spectrum within the RPA between the two theories.

This paper is structured as follows: We begin in Sec.~\ref{sec:motivation} by introducing the Son-Dirac theory and its symmetries, and we discuss aspects of the non-Galilean invariant response that motivate the present study. Section~\ref{sec:rpa} contains a field-theoretical derivation of the response in the half-filled Landau level using the RPA. The results of this calculation are presented in Sec.~\ref{sec:results}, which discusses in particular the response at long wavelengths. The paper is concluded by a summary in Sec.~\ref{eq:summary}. There are two Appendixes: Appendix~\ref{app:hlr} collects results for the response within a modified HLR theory projected to the LLL for reference. Appendix~\ref{app:B} computes the noninteracting response functions of two-dimensional Dirac fermions used in the main text, which are linked to the electron response via the RPA.

\section{Definitions and Motivation}\label{sec:motivation}

The theory proposed by Son is a Chern-Simons theory for Dirac fermions with Lagrangian density~\cite{son15}
\begin{align}
{\cal L}_{\rm SD} &= \psi^\dagger \bigl[ (i \hbar \partial_t + e a_0) + v_F \sigma^i \bigl(i \hbar \partial_i + e a_i\bigr)  \, \bigr] \psi \nonumber \\*
&\qquad - \frac{e}{2 \phi_0} \, \varepsilon^{\mu\nu\rho} A_\mu \partial_\nu a_\rho + \frac{e}{4 \phi_0} \, \varepsilon^{\mu\nu\rho} A_\mu \partial_\nu A_\rho , \label{eq:SDlagrangian}
\end{align}
where $\phi_0 = 2\pi\hbar/e$, with $e$ the electron charge, $\psi$ is the two-component Dirac fermion field, $a^\mu$ is the Chern-Simons gauge field, $A^\mu$ is the external vector potential, and the Fermi velocity $v_F$ is an effective parameter that sets the strength of the Coulomb interaction. A summation convention is implied in this paper, where Greek indices run over $\mu=0,1,2$, with $0$ (or $t$) a temporal index and the latin index $i=1,2$ (or $x,y$) a space index, and $\varepsilon^{\mu\nu\rho}$ is the total antisymmetric tensor with $\varepsilon^{012} = 1$. Particle-hole symmetry is realized as a combination of time-reversal and charge conjugation, with a transformation $(A_0'(t, {\bf x}), A_i'(t, {\bf x})) = (- A_0(- t, {\bf x}), A_i(- t, {\bf x}))$, $(a_0'(t, {\bf x}),a_i'(t, {\bf x})) = (a_0(- t, {\bf x}),- a_i(- t, {\bf x}))$, and $\psi'(t, {\bf x}) = - i \sigma_2 \psi(- t, {\bf x})$~\cite{son15}. Note that the particle-hole symmetry operation on this level is implemented as an antiunitary transformation (for a review, see~\cite{zirnbauer21}). Different from HLR theory, the composite Dirac fermions do not couple directly to the external gauge field but only indirectly through the mixed Chern-Simons term [the first term in the second line of Eq.~\eqref{eq:SDlagrangian}]. In the mean-field approximation at half-filling, where the electron density is $j^0 =1/4\pi\ell_B^2$ with $\ell_B = \sqrt{\hbar/eB}$ the magnetic length, the composite Dirac fermions experience no effective Chern-Simons magnetic field $\langle b \rangle = \langle \varepsilon^{ij} \partial_i a_j \rangle = 0$. On that level, composite fermions have a valence and a conduction band with linear dispersion $\pm \hbar v_F q$, and (barring spontaneous symmetry breaking~\cite{kamburov14,barkeshli14,mitra19}) they form a Fermi sea with Fermi momentum $k_F = 1/\ell_B$ and a Fermi energy that is detuned from the Dirac point by $E_F = \hbar v_F/\ell_B$. Note that Eq.~\eqref{eq:SDlagrangian} is an effective field theory, which may contain additional terms that, for example, involve higher derivatives or powers of the fields.

The Son-Dirac theory~\eqref{eq:SDlagrangian} is not invariant under Galilei transformation. A modified version of the Son-Dirac theory (which we shall refer to as the modified Son-Dirac theory) is brought to Galilean-invariant form by coupling the composite Dirac fermions directly to the external electric field $E_i$ by adding a dipole term~\cite{son18}
\begin{align}
{\cal L}_D &= {\bf d} \cdot {\bf E} \label{eq:dipole}
\end{align}
 to the action~\eqref{eq:SDlagrangian}  with $E_i = \partial_i A_0 - \partial_0 A_i$ and a composite-fermion dipole moment
\begin{align}
d^i &= \frac{\varepsilon^{ji}}{2 B} \bigl[ \psi^\dagger \bigl(i \hbar \partial_j + e a_j\bigr) \psi + \bigl(- i \hbar \partial_j + e a_j\bigr) \psi^\dagger \psi\bigr] . \label{eq:dipolemoment}
\end{align}
A Galilei transformation to a moving inertial frame with coordinates ${\bf x}' = {\bf x} - {\bf V} t$ is then implemented by $A_0'(t',{\bf x}') = A_0(t,{\bf x}) + V^i A_i(t,{\bf x})$, $A_i'(t',{\bf x}') = A_i(t,{\bf x})$ (such that $\varepsilon^{ij} E_i' = \varepsilon^{ij} E_i + V^j B$ and $B' = B$) and $\psi'(t', {\bf x}') = \psi(t, {\bf x})$ (the Chern-Simons field $a_\mu$ transforms in the same way as the field $A_\mu$). Intuitively, the dipole moment \mbox{${\bf d} = e \ell_B^2 \hat{\bf z} \times {\bf k}$} of a composite fermion with momentum $k$ arises due to a separation of the electron and vortex position by $\ell_B^2 k$ and is a fundamental feature of the half-filled Landau level~\cite{read94}.

\begin{figure}[t]
\scalebox{0.82}{\includegraphics{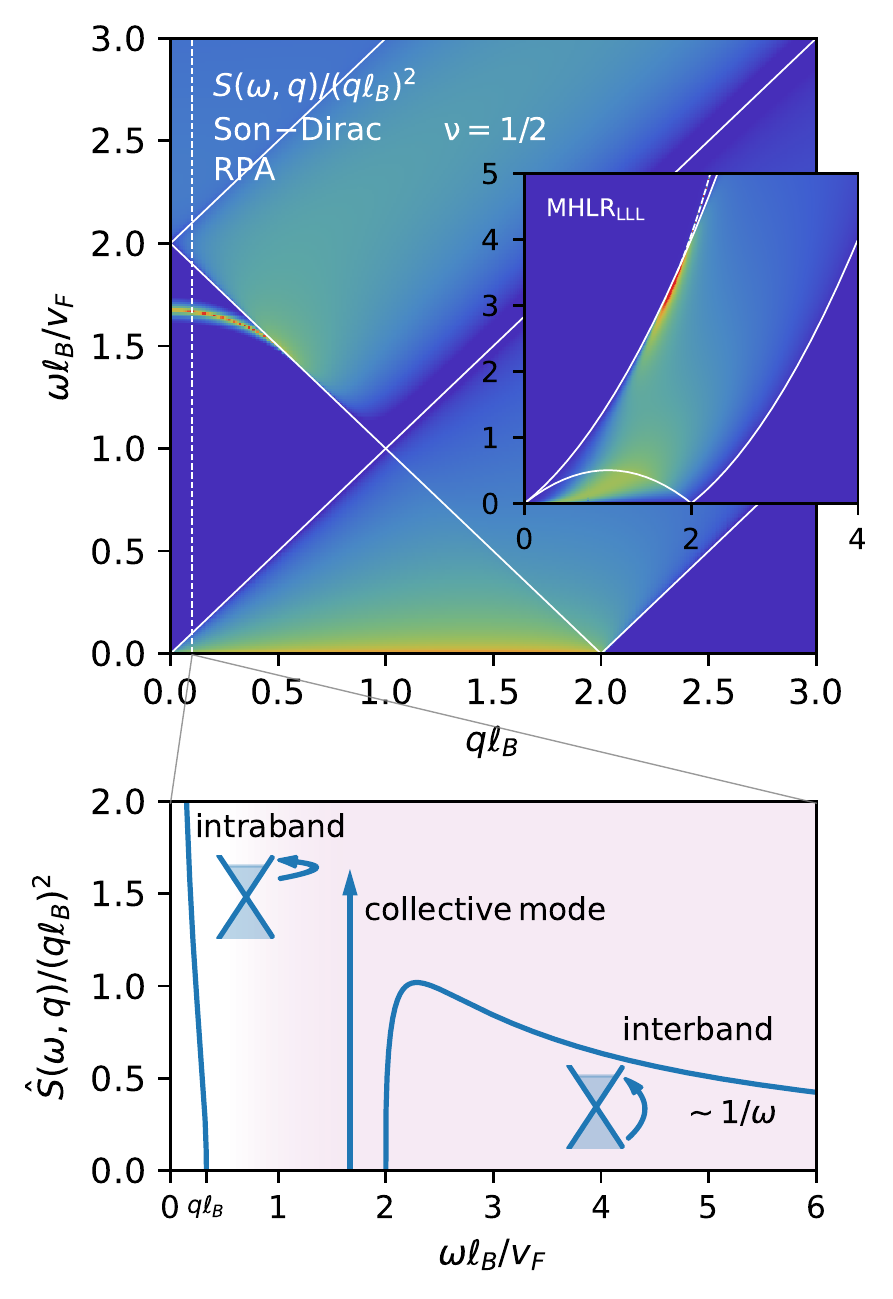}}
\caption{Dynamic structure factor of the unmodified Son-Dirac theory at half-filling as predicted by the RPA. There is a Fermi-liquid intraband contribution at low frequencies $\omega < v_F q$, but also an additional high-frequency part caused by interband transition of the composite Dirac fermions. A collective mode at high frequencies and long wavelengths is also apparent. Bottom figure: Dimensionless dynamic structure factor as a function of frequency in the long-wavelength limit (marked by a dashed line in the full figure). The high-frequency response (red shaded area) is spurious. Inset: Dynamic structure factor as predicted by the modified HLR theory, which shows no interband transitions.}
\label{fig:1}
\end{figure}

\begin{figure}[t]
\scalebox{0.845}{\includegraphics{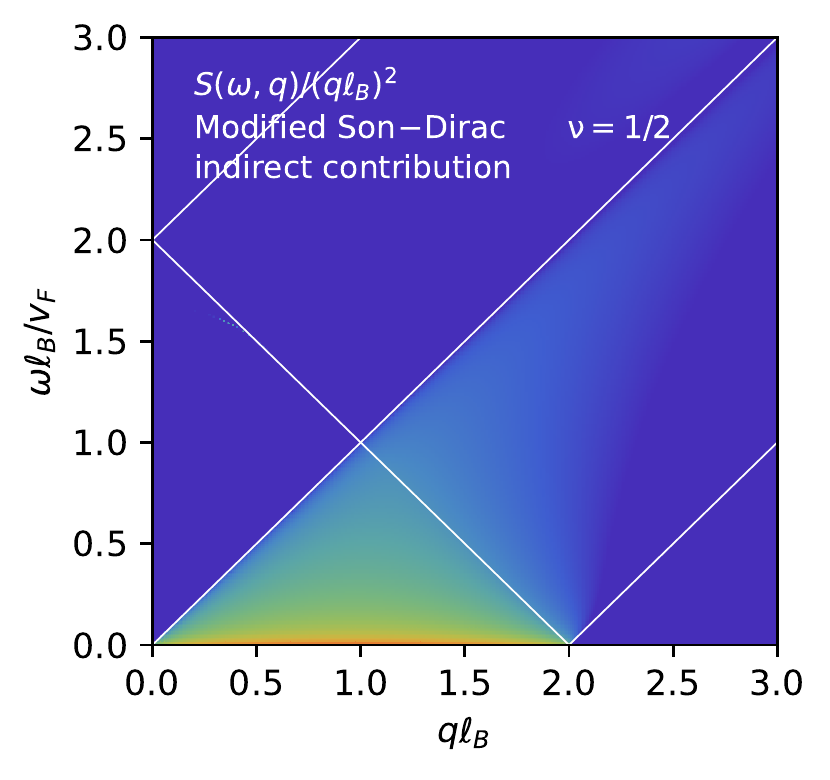}}
\caption{Dynamic structure factor computed within the RPA for the modified Galilean-invariant Son-Dirac theory, which includes the dipole correction. Compared to Fig.~\ref{fig:1}, interband transitions and the collective mode are now strongly suppressed. The same color-coding is used in both figures for the intensity of the response.}
\label{fig:2}
\end{figure}

To motivate the current investigation, consider the density response of the theory~\eqref{eq:SDlagrangian} without including the dipole term that ensures Galilean invariance. In this case, the frequency- and momentum-dependent density response $\Pi^{00}$ within the RPA is given by~\cite{son15}
\begin{align}
\Pi^{00}(\omega, q) &= \biggl(\frac{q}{4\pi\hbar}\biggr)^2  \frac{1}{[K^{xx}(\omega, x)]^*} , \label{eq:SDRPA1}
\end{align}
where $K^{xx}$ is the transverse noninteracting current response function of Dirac fermions (a discussion and derivation of this result is given in the remainder of the paper). Figure~\ref{fig:1} shows a density plot of the corresponding dynamic structure factor
\begin{align}
S(\omega, q) &= \frac{1}{\pi} {\rm Im} \, \Pi^{00}(\omega, q) ,
\end{align}
where the long-wavelength region $q\to 0$ as a function of frequency is shown at the bottom of Fig.~\ref{fig:1}. The excitation spectrum takes a form that is typical for Dirac fermions, with incoherent spectral weight in the low-frequency region $\omega < v_F q$ due to intraband excitations [i.e., particle-hole excitations within the conduction band], and further weight at larger frequencies due to interband transitions between the valence and the conduction band of the Dirac fermions. The weight of both the intra- and interband excitation at long wavelengths is of order ${\cal O}(q^2)$, and the interband contribution at high frequencies decays as ${\cal O}(q^2/\omega)$. There is no continuous spectral weight in a wedge $|\omega - \omega_c| < v_F q$ due to phase-space restrictions. In this region, the RPA predicts a well-defined collective mode at long wavelengths $q\ell_B/\hbar \lesssim 0.5$ starting at a frequency $ \Omega \approx 1.667 \omega_c$ with a residue of order ${\cal O}(q^2)$. For comparison, we show as an inset in Fig.~\ref{fig:1} the corresponding result for the dynamic structure factor of a modified version of HLR theory that is projected to the lowest Landau level (cf. Appendix~\ref{app:hlr} for a detailed discussion). The HLR theory shows a (presumably spurious) collective mode that decouples from the Fermi-liquid continuum at finite wave vectors, but crucially, the response contains no interband transitions.

While the high-frequency response shown in Fig.~\ref{fig:1} is typical for Dirac fermions, as discussed, it is unexpected for a theory of electrons in the half-filled Landau level, for which a Dirac cone does not exist. Indeed, at least within HLR theory (not projected to the LLL), the only large-frequency response of electrons in a magnetic field is associated with transitions between electron Landau levels, with a typical energy scale of the order of the cyclotron frequency~\cite{girvin86}, which is much larger than the scales considered here and projected out in a theory restricted to the LLL. In addition, the only collective finite-frequency mode is the magnetoplasmon, the frequency of which is fixed by Kohn's theorem again at the cyclotron frequency~\cite{kohn61}. This magnetoplasmon exhausts the $f$-sum rule
\begin{align}
f(q) &= \int_0^\infty \frac{d\omega}{\pi} \, \omega \, {\rm Im} \, \Pi^{00}(\omega, q) \label{eq:fsumrule}
\end{align}
at long wavelengths and is thus the only mode that contributes at ${\cal O}(q^2)$ to the dynamic structure factor. When restricted to the lowest Landau level, the $f$-sum rule is of order ${\cal O}(q^4)$~\cite{girvin86}, which is in contrast to the weight of the Dirac interband spectrum in Fig.~\ref{fig:1}. Even if this contribution had the correct weight, the \mbox{$f$-sum} rule would diverge linearly on account of the $1/\omega$ high-frequency tail (cf.~Fig.~\ref{fig:1}). Note that spurious interband excitations and a collective mode also appear in the transverse current response, the spectral function of which takes a similar structure to that of the density response in Fig.~\ref{fig:1}.

The aim of this paper is to demonstrate that the composite Dirac fermion theory~\eqref{eq:SDlagrangian} can be extended to give rise to an RPA response valid at all frequencies, at least to leading order in the wave number. We show that this is the case for the modified Son-Dirac theory with the dipole term~\eqref{eq:dipole} that restores Galilean invariance. As an illustration and preempting a main result derived in the remainder of this paper, Fig.~\ref{fig:2} shows the response for the Galilean-invariant theory, where the spurious high-frequency response is indeed strongly suppressed, thus reproducing this feature of HLR theory. Previous literature focusses on the non-Galilean invariant theory~\cite{son15} or on the semiclassical limit in the vicinity of the half-filled state $\nu = 1/2 \pm 1/2N$~\cite{nguyen18}, for which the response is accurate in a $1/N$-expansion for small wave numbers ${\cal O}(1/\ell_BN)$ and energies ${\cal O}(\hbar v_F/\ell_BN)$. Such a rigorous power-counting argument does not apply at half-filling, and a full effective theory of the half-filled Landau level will include additional terms beyond~\eqref{eq:SDlagrangian} and~\eqref{eq:dipole}. Here, a consistent theory that eliminates high-frequency modes will constrain the effective theory beyond the leading \mbox{order}. 

\section{RPA for the half-filled Landau level}\label{sec:rpa}

In this section, we consider the functional integral of the Son-Dirac partition function and derive the effective action to second order in fluctuations of the gauge fields around the mean-field result, which gives the RPA response functions. To this end, we first summarize in Sec.~\ref{sec:constraints} the various constraints as well as the conjugate density and current corresponding to the Lagrangian~\eqref{eq:SDlagrangian}. In Sec.~\ref{sec:rpaderivation}, we then consider the Euclidean path integral and derive the effective action.

\subsection{Constraints and conjugate fields}\label{sec:constraints}

The action~\eqref{eq:SDlagrangian} is linear in the Chern-Simons field $a_0$, which is a Lagrange multiplier that enforces a constraint on the composite fermion density
\begin{align}
J^0 = \psi^\dagger \psi &= \frac{B}{2 \phi_0} . \label{eq:constr1}
\end{align}
The external magnetic field thus sets the density of composite fermions. This is different from HLR theory, where the electron density (which in HLR is equal to the composite fermion density) constrains the Chern-Simons magnetic field $b$~~\cite{hlr93}. Likewise, the spatial components $a_i$ enforce a constraint on the composite fermion current~\cite{son15,prabhu17,son18}
\begin{align}
J^i = v_F \psi^\dagger \sigma^i \psi &= - \frac{1}{2 \phi_0} \, \varepsilon^{ij} E_j + \frac{1}{2 \phi_0} \, \varepsilon^{ij} E_j = 0 .  \label{eq:constr2}
\end{align}
The second term, which cancels the first contribution, arises from the dipole correction~\eqref{eq:dipole} in conjunction with the constraint~\eqref{eq:constr1}. Galilean invariance thus induces a backflow correction such that the composite Dirac fermion current does not respond to the electric field.

Unlike in HLR theory, the composite particle density $J^0$ and current $J^i$ in the Son-Dirac theory are not equal to the conserved electron density and current. The electron density, which is conjugate to the field $eA_0$, is given by~\cite{son18}
\begin{align}
j^0 &= \frac{1}{2\phi_0} (B - b) - \partial_i d^i , \label{eq:constraint}
\end{align}
where $b = \varepsilon^{ij} \partial_i a_j$, and $d^i$ is the dipole moment defined in Eq.~\eqref{eq:dipolemoment}. The last term is just the polarization charge of dipoles with dipole moment $d^i$. Using the definition in Eq.~\eqref{eq:dipolemoment}, this expression can be cast in a different form,
\begin{align}
j^0 &= \frac{B}{2\phi_0} - \varepsilon^{ij} \partial_i \Bigl(\frac{- i \hbar}{2 B} (\psi^\dagger \overset\leftrightarrow{\partial}_j \psi)\Bigr) \label{eq:vorticity} .
\end{align}
Identifying the external magnetic field $B$ with the density of composite fermions, Eq.~\eqref{eq:constr1}, we see that density fluctuations are linked to the vorticity of the Dirac field. Likewise, the current conjugate to the external field $eA_i$ is
\begin{align}
j^i &= \frac{\varepsilon^{ij}}{2 \phi_0} \, (E_j - e_j) + \partial_0 d^i + \varepsilon^{ji} \partial_j m
\label{eq:constraint2}
\end{align}
with $e_j = \partial_j a_0 - \partial_0 a_j$ and $m = \frac{\varepsilon^{jk} E_k}{2 B^2} i \hbar (\psi^\dagger \overset\leftrightarrow{\partial}_j \psi)
$. The last two terms again follow from the dipole term~\eqref{eq:dipole} that restores Galilean invariance. The second-to-last term is the contribution induced by electric dipoles with dipole moment $d^i$, and the last term --- which arises from a variation of the magnetic field in the denominator of Eq.~\eqref{eq:dipolemoment} --- is characteristic for the current induced by magnetic dipoles with magnetization ${\bf M} = m \hat{\bf e}_z$. On a mean field level, the definition of the density~\eqref{eq:constraint} along with
\begin{align}
\nu &= \frac{\phi_0}{B} \langle j^0 \rangle = \frac{1}{2}
\end{align}
implies that the expectation value of the Chern-Simons magnetic field $\langle b \rangle$ is zero, such that the composite Dirac fermions do not experience an effective magnetic field. As we consider an isotropic system, the corrections to the particle density and current arising from the dipole term~\eqref{eq:dipole} do not contribute to the mean-field result. They will, however, change the fluctuations (which now couple directly to the external gauge field through the dipole term) and thus the response functions.

\subsection{Random phase approximation}\label{sec:rpaderivation}

In this section, we derive the linear response function for the density and current of the Son-Dirac theory on the level of the RPA. To this end, we consider the Euclidean path integral and expand around the mean-field saddle point up to second order in the external gauge fields. The kernel of this expansion is related to the response functions by analytical continuation. As discussed in the Introduction, the advantage of the RPA is that the response functions of the interacting theory are expressed in terms of the free noninteracting response functions of 2D Dirac fermions, which can be computed in closed analytical form (cf. Appendix~\ref{app:B}).

The starting point is the Euclidean partition function
\begin{align}
{\cal Z} [A] &= \int {\cal D}[\psi^\dagger, \psi, a_\mu] \, e^{- S_{\rm E}[\psi^\dagger, \psi, a_\mu, A_\mu]} , \label{eq:partitionfunction}
\end{align}
where $S_{\rm E}$ is the Euclidean action corresponding to the Lagrangian~\eqref{eq:SDlagrangian} (i.e., obtained after a Wick rotation to imaginary time $t = - i \tau \hbar$),
\begin{align}
{\cal L}_{SD}^E &= \psi^\dagger \bigl[ (\partial_\tau - e a_0) + v_F \sigma^i \bigl(- i \hbar \partial_i - e a_i\bigr)  \, \bigr] \psi - d^i E_i \nonumber \\*
&\quad + \frac{e}{2 \phi_0} \, \varepsilon^{\mu\nu\rho} A_\mu \partial_{E\nu} a_\rho - \frac{e}{4 \phi_0} \, \varepsilon^{\mu\nu\rho} A_\mu \partial_{E\nu} A_\rho , 
\end{align} 
where the derivative is now $\partial_{E\mu} = (\tfrac{i}{\hbar} \partial_\tau, \partial_{\bf x})$. In the following, we denote by $\bar{A}_\mu$ a background-field configuration with constant magnetic field $\bar{B}$ such that $B = \bar{B} + \delta B$ and $E_i = \delta E_i$, i.e., we split off the gauge field fluctuations as
\begin{align}
A_\mu &= \bar{A}_\mu + \delta A_\mu .
\end{align}
In addition, we have \mbox{$\bar{a}_i = 0$} in the half-filled Landau level, and we split off fluctuations in the Chern-Simons field as $a_0 = \bar{a}_0 + \delta a_0$ and $a_i = \delta a_i$. 
 Within linear response, we have
\begin{align}
\langle j^\mu(x) \rangle &= \int dy \, \Pi^{\mu\nu}(x,y) \, e \delta A_\nu(y) ,
\end{align}
where the density with $\mu=0$ is given in Eq.~\eqref{eq:constraint} and the current with $\mu=i$ is given in Eq.~\eqref{eq:constraint2}, and we use a three-vector notation $x = (\tau,{\bf x})$ for the coordinates. The response kernel is given by
\begin{align}
\Pi^{\mu\nu}(x,y) &= 
- \frac{\delta^2 \ln {\cal Z}[\delta A]}{\delta (e \delta A_\mu(x)) \delta (e \delta A_\nu(y))} \biggr|_{ A_\mu=\bar{A}_\mu} . \label{eq:responsefunctions}
\end{align}
We shall work in Coulomb gauge where $\nabla \cdot {\bf A} = 0$. In Fourier space with an external momentum oriented along the $y$-axis, ${\bf q} = (0,|{\bf q}|)$, this implies $A_y = 0$ [note that we will continue to use Greek indices for the summation over the $0$ and $x$ component]. In this convention, $\Pi^{00}$ denotes the density response function, $\Pi^{xx}$ denotes the transverse current response functions, and $\Pi^{0x}$ denotes the mixed density-current response. 

\begin{figure}[t]
\subfigure[]{\raisebox{0.3cm}{\scalebox{0.59}{\includegraphics{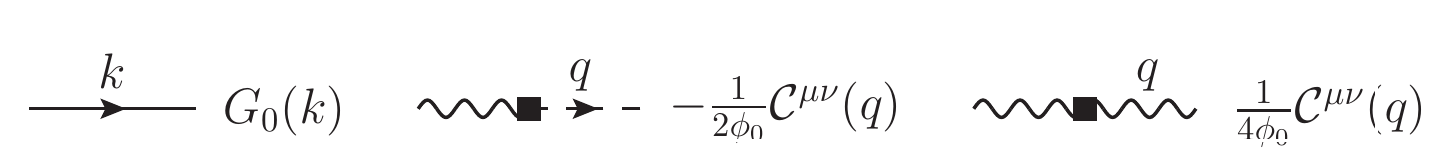}}}}
\subfigure[]{\scalebox{0.5}{\includegraphics{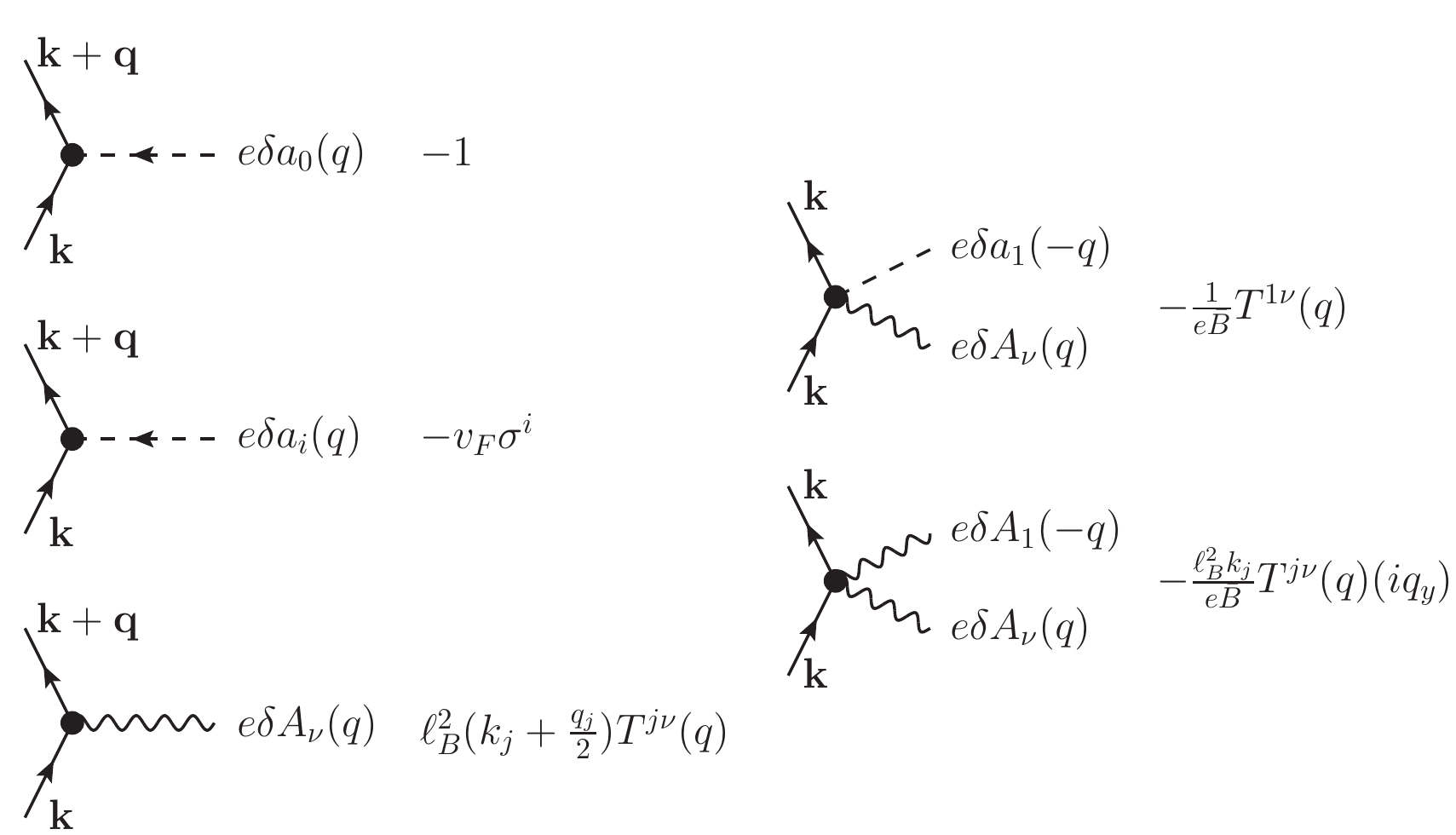}}}
\caption{(a) Feynman rules for the Dirac propagator and the Chern-Simons terms. The Dirac field is indicated by a continuous line, the Chern-Simons gauge field is shown by a dashed line, and the external vector potential is shown by a wavy line. (b) Feynman rules for the vertex terms that couple the composite Dirac field to the Chern-Simons gauge field and the external vector potential. Terms on the left-hand side contribute at leading linear order in the gauge field fluctuations, and terms on the right-hand side at second order. For the latter, we only show momentum configurations that will contribute a diamagnetic term to the response.}
\label{fig:vertices}
\end{figure}

The Son-Dirac action is quadratic in the composite fermion fields, such that the Grassmann integral can be performed directly. This gives an effective Euclidean action
\begin{align}
S_{\rm eff} &= - \, {\rm Tr} \ln [- G^{-1}] + \frac{e}{\phi_0} \int_q \Big\{ 
A^T(-q) \frac{{\cal C}(q)}{2} a(q)\nonumber \\*
& \quad - A^T(-q) \frac{{\cal C}(q)}{4} A(q) \Bigr\} ,\label{eq:eff1}
\end{align}
where $G^{-1}$ is the inverse Green's function, and we define $A = (A_0, A_1)$ as well as
\begin{align}
{\cal C}(q) &=
\begin{pmatrix}
0 & - i q_y \\ i q_y & 0
\end{pmatrix} .
\end{align}
In position space, the Green's function can be written in a form that separates out the fluctuations,
\begin{align}
G^{-1}(x) &= G_0^{-1}(x) - V(x) , \label{eq:greens}
\end{align}
where $G_0$ is the bare propagator with Fourier transform
\begin{align}
G_0^{-1}(q) &= i \hbar \omega - ( v_F \hbar{\bf q}\cdot \boldsymbol{\sigma} - \bar{a}_0) . \label{eq:G0}
\end{align}
This is the free propagator of two-dimensional Dirac electrons, where the mean-field contribution $\bar{a}_0$ acts as a chemical potential. 
The Fourier transform of the fluctuation terms reads
\begin{align}
V^{(1)}(k|q) &= - e \delta a_0(q) - v_F \sigma^i (e \delta a_i(q)) \nonumber \\*
& \quad + \frac{(\hbar {\bf k} + \hbar {\bf q}/2) \times \delta {\bf E}(q)}{\bar{B}} 
\label{eq:V1}
\end{align}
to leading order in the fluctuations, and
\begin{align}
V^{(2)}(k|q,q') &= \frac{\delta {\bf E}(q) \times e \delta {\bf a}(q')}{\bar{B}} \nonumber \\*
& \quad - \frac{(\hbar{\bf k} +  \hbar({\bf q} +  {\bf q}')/2) \times \delta {\bf E}(q)}{\bar{B}^2}  \delta B(q')
\label{eq:V2}
\end{align}
to second order, where the electric field is in the plane (we neglect fluctuations beyond quadratic order, which will not contribute to the RPA response). Pictographically, the fluctuations $V^{(1)}$ and $V^{(2)}$ describe vertex terms that couple the Dirac electrons to the external gauge field and the Chern-Simons gauge field, respectively. These vertices are shown in Fig.~\ref{fig:vertices}, where it is convenient to introduce the conversion matrix
\begin{align}
T^{i\nu}(q) &=
\begin{pmatrix}
i q_y & 0 \\ 
0 & \omega
\end{pmatrix} ,
\end{align}
which maps the external vector potential to the electric field fluctuation, $\varepsilon^{ij} \delta E_j(q) = [T(q)]^{i \nu} \delta A_\nu(q)$. In Fig.~\ref{fig:vertices}, continuous lines denote the Dirac field, the wavy line denotes the external field, and the dashed line denotes the Chern-Simons field. Only the second vertex contains Pauli matrices, while the remaining terms are diagonal in spinor space. The last term in Eq.~\eqref{eq:V1} and both terms in~\eqref{eq:V2} are due to the dipole term. The remaining Feynman rules are then as usual, where one imposes momentum and energy conservation at each vertex and integrates over each undetermined loop momentum with measure \mbox{$\int_k = \int d(\hbar\omega) d{\bf k}/(2\pi)^3$}.

\begin{figure}[t]
\subfigure[]{\raisebox{0.2cm}{\scalebox{0.5}{\includegraphics{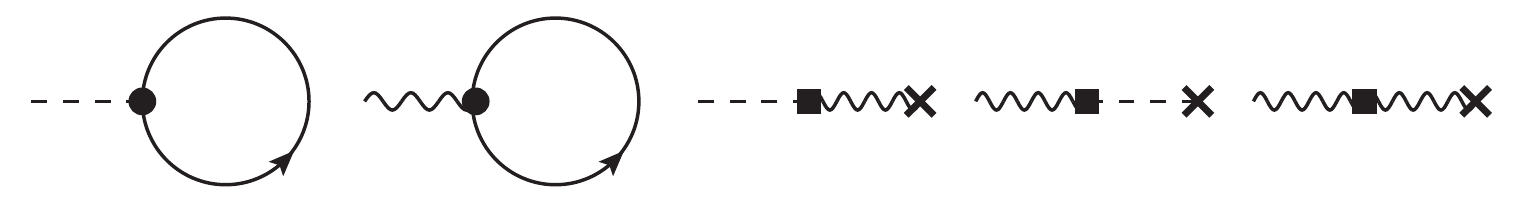}}}}\\[3ex]
\subfigure[]{\scalebox{0.5}{\includegraphics{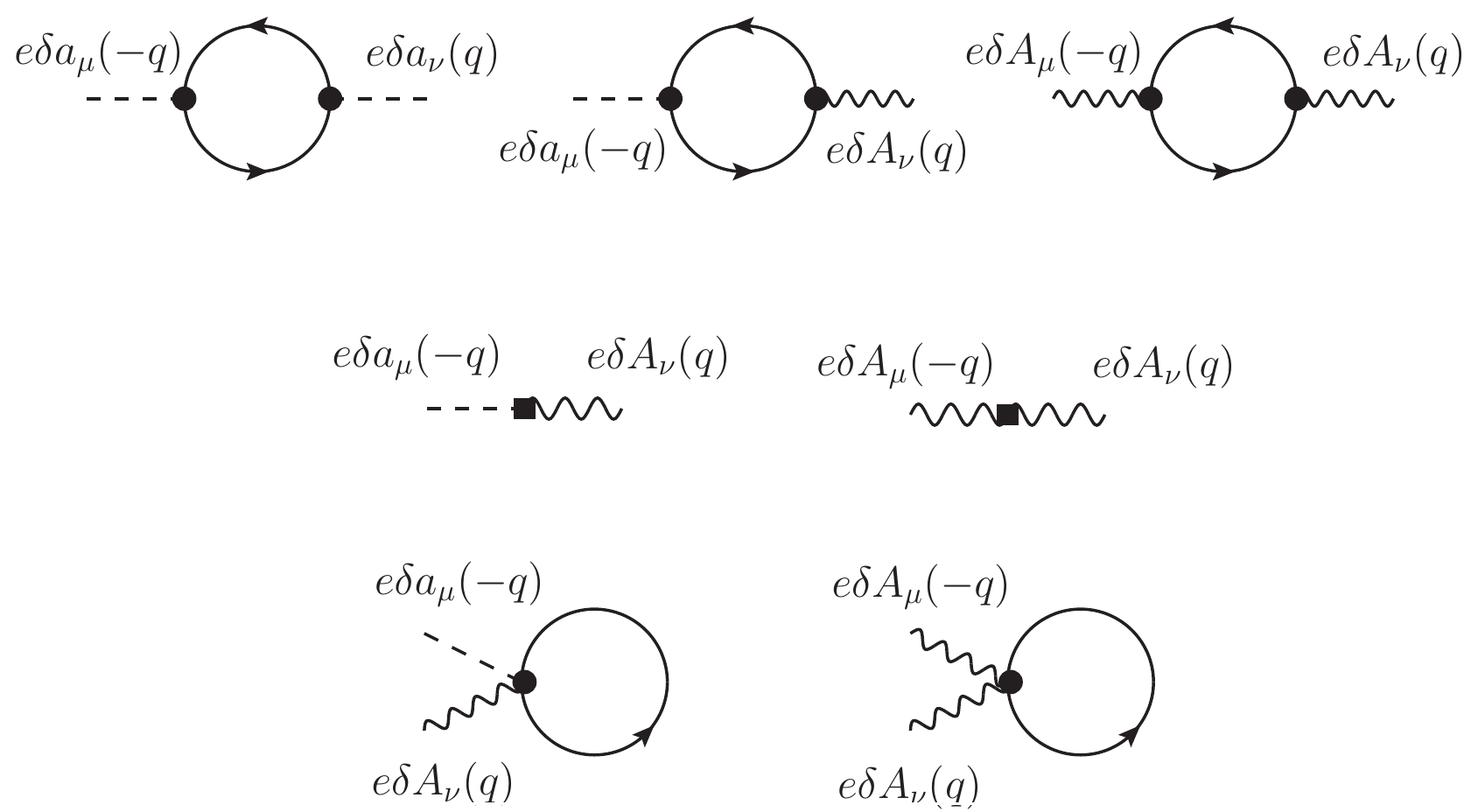}}}
\caption{Diagrams contributing to the effective action up to second order in the gauge field fluctuations. (a) Contributions at linear order. The first two diagrams represent tadpole diagrams, some of which are zero. (b) Contributions at quadratic order in the gauge fields. The first line shows paramagnetic response functions, the second line shows the Chern-Simons terms, and the third line shows diamagnetic terms.}
\label{fig:feynman}
\end{figure}

In the following, we use the decomposition~\eqref{eq:greens} to expand the trace of the logarithm~\eqref{eq:eff1} to second order in the field fluctuations $\delta a_\mu$ and $\delta A_\mu$. At the leading linear order, the effective action reads (omitting terms that evaluate to zero)
\begin{align}
&S_{\rm eff}^{(1)} = - \int_k \, {\rm tr} [G_0(k)] \delta a_0(0) + \frac{\bar{B}}{2\phi_0} \delta a_0(0) - \frac{\bar{B}}{2 \phi_0} \delta A_0(0) , \label{eq:Seff1}
\end{align}
where the trace runs over the spinor indices. The corresponding Feynman diagrams are shown in Fig.~\ref{fig:feynman}(a). The first term in Eq.~\eqref{eq:Seff1} follows from the leading-order expansion of the logarithm in Eq.~\eqref{eq:eff1} [diagrammatically, this is the first tadpole diagram in Fig.~\ref{fig:feynman}(a), with the second evaluating to zero], the second line is the mixed Chern-Simons term, and the last line is the Chern-Simons term for the external gauge field. 
Evaluated at the saddle point, this contribution has to vanish. Indeed, varying with respect to $\delta a_0$ and $\delta a_i$, we reproduce the constraints~\eqref{eq:constr1} and~\eqref{eq:constr2} to leading order,
\begin{align}
\langle J^0 \rangle &= \int_k \, {\rm tr} [G_0(k)] = \frac{\bar{B}}{2\phi_0} , \label{eq:conmf} \\
\langle J^i \rangle &= v_F \int_k \, {\rm tr} [G_0(k) \sigma^i] = 0 \label{eq:conmf2} .
\end{align}
The value of the mean-field contribution $\bar{a}_0$ is adjusted to set the density~\eqref{eq:conmf}. Relation~\eqref{eq:conmf2} is satisfied by dint of the symmetry properties of free Dirac fermions [the corresponding term is omitted in Eq.~\eqref{eq:Seff1}].  
In addition, the first variation with respect to $\delta A_\mu$ gives a mean-field density of $n = \bar{B}/2\phi_0$ and a vanishing mean-field current, as expected for the half-filled Landau level.

To second order in the field fluctuations, the effective action is (again omitting terms that evaluate to zero)
\begin{align}
S_{\rm eff}^{(2)} &= \frac{1}{2} \int_q
\Bigl\{
e^2 \delta a^T(-q) \, {\cal K}(q) \, \delta a(q) \nonumber \\
&\quad - \frac{e}{2 \phi_0} \, \delta a^T(-q) \, \bigl[{\cal R}(q) {\cal T}(q)\bigr] \, \delta A(q) \nonumber \\
&\quad - \frac{e}{2 \phi_0} \, \delta A^T(-q) \, \bigl[{\cal T}^T(-q) {\cal R}^T(-q)\bigr] \, \delta a(q) \nonumber \\
&\quad + e^2 \delta A^T(-q) \, \bigl[{\cal T}^T(-q) \Delta(q) {\cal T}(q)\bigr] \, \delta A(q) 
\Bigl\} \nonumber \\
& \quad - \frac{e}{2 \phi_0} \, \int_q \delta a^T(-q) \, {\cal C}(q) \delta A(q)  \nonumber \\
& \quad + \frac{e}{4 \phi_0} \, \int_q \delta A^T(-q) \, {\cal C}(q) \delta A(q)  \nonumber \\
& \quad+ \frac{1}{e \bar{B}} \int_k {\rm tr} [G_0(k)] \int_q \, e \delta a_1(-q) \, [T(q)]^{1\nu} \, e A_\nu(q) . \label{eq:Seff2}
\end{align}
Diagrammatically, the various contributions are shown in Fig.~\ref{fig:feynman}(b). The first four terms in Eq.~\eqref{eq:Seff2} in curly brackets [first line in Fig.~\ref{fig:feynman}(b)] are contributions to a paramagnetic response, where the Dirac fermions are integrated out at the one-loop level. They are defined as
\begin{widetext}
\begin{align}
{\cal K}^{\mu\nu}(q) &= \int_k \, {\rm tr} 
\begin{pmatrix}
G_0(k) G_0(k+q) &
 v_F G_0(k) \sigma^1 G_0(k+q) 
\\
 v_F \sigma^1 G_0(k) G_0(k+q) & v_F^2 \sigma^1G_0(k) \sigma^1 G_0(k+q)
\\
\end{pmatrix} \label{eq:defK} \\[2ex]
{\cal R}^{\nu i}(q) &= 4 \pi \ell_B^2 
 \int_k \, {\rm tr} \begin{pmatrix}
- {\hbar k_x} G_0(k) G_0(k+q) & ({\hbar k_y+\hbar q_y/2}) G_0(k) G_0(k+q) \\[1ex]
v_F {\hbar k_x} G_0(k) \sigma^1 G_0(k+q) & - v_F ({\hbar k_y+\hbar q_y/2}) G_0(k) \sigma^1 G_0(k+q) 
\end{pmatrix} \\[2ex]
\Delta^{ij}(x,y) &= \ell_B^4
\int_k \, {\rm tr} \begin{pmatrix}
{k_x^2} G_0(k) G_0(k+q) & { k_x ( k_y +  q_y/2)} G_0(k) G_0(k+q) \\[1ex]
{ k_x ( k_y +  q_y/2)} G_0(k) G_0(k+q) & {( k_y +  q_y/2)^2}  G_0(k) G_0(k+q) 
\end{pmatrix} . \label{eq:defDelta}
\end{align}
\end{widetext}
These ``Dirac Lindhard functions'' can be evaluated in closed analytical form, which is done in Appendix~\ref{app:B}. The mixed response function ${\cal R}(q)$ as well as the direct dipole response to the external field $\Delta(q)$ are due to the dipole terms in Eqs.~\eqref{eq:constraint} and~\eqref{eq:constraint2}. Returning to Eq.~\eqref{eq:Seff2}, the next two terms in the second and third line [second line in Fig.~\ref{fig:feynman}(b)] arise from the Chern-Simons terms. Finally, the last term [third line in Fig.~\ref{fig:feynman}(b)] is a diamagnetic contribution. This term cancels with the $x0$-contribution of the mixed Chern-Simons term two lines prior. We express the sum of these two terms using a new conversion matrix
\begin{align}
\tilde{C}(q) &=
\begin{pmatrix}
0 & - i q_y/\omega \\
0 & 0
\end{pmatrix}
\end{align}
such that the mixed Chern-Simons term reads $- \frac{e}{2 \phi_0} \, \int_q \delta a^T(-q) \, \tilde{\cal C}(q) T \delta A(q)$. A similar cancellation between the dipole correction and the Chern-Simons term was noted in Ref.~\cite{prabhu17}.

The RPA consists of bringing the effective action to quadratic form in the Chern-Simons fluctuation $\delta a_\mu$, in which case the Gaussian path integral over that field decouples. This is accomplished by shifting
\begin{align}
e \delta a(q) &\to e \delta a(q) -  \frac{1}{2 \phi_0} \, [{\cal K}^{-1}] ({\cal R} + \tilde{\cal C}) T \delta A(q) ,
\end{align}
which gives the full effective action at second order in the external field:
\begin{align}
S_{\rm eff} &= - \frac{1}{2} \int_q \, [T e \delta A(-q)]^T \bigl\{ \Delta + \Bigl(\frac{1}{4 \pi \hbar}\Bigr)^2  \nonumber \\*
& \quad  \times ( {\cal R} + \tilde{\cal C}(-q))^T {\cal K}^{-1}(-q) ({\cal R} + \tilde{\cal C}(q)) \bigr\} [T e \delta A(q)] \nonumber \\*
&\quad+ \frac{e}{4 \phi_0} \, \int_q \delta A^T(-q) {\cal C} \delta A(q) .
\label{eq:effective action}
\end{align}
Using Eq.~\eqref{eq:responsefunctions}, this effective action determines the linear response functions. The retarded response functions are then obtained by analytic continuation from Euclidean imaginary frequencies to real frequencies, $i\omega \to \omega + i0$. The results of this calculation are discussed in the next section.

\section{Results}\label{sec:results}

In this section, we discuss the results for the density response function, the transverse current response function, and the Hall response function. Using the result~\eqref{eq:effective action} to compute the response function~\eqref{eq:responsefunctions} gives
\begin{align}
&\Pi^{\mu\nu}(q) = \frac{1}{4 \pi \hbar} {\cal C}^{\mu\nu} - [T(-q) \Delta T(q)]^{\mu\nu} \nonumber \\*
& + \Bigl(\frac{1}{4 \pi \hbar}\Bigr)^2 \bigl[ T(-q) ({\cal R} + \tilde{\cal C})^T {\cal K}^{-1} ({\cal R} + \tilde{\cal C}) T(q)\bigr]^{\mu\nu} .
\end{align}
There are three separate contributions to this function: The first term arises from the $AdA$ term; the second one is a direct response of noninteracting dipoles that couple directly to the external field; and the third term is an indirect response, where composite fermions couple to the external probe through the Chern-Simons field. Different from the non-Galilean invariant theory, this indirect coupling is no longer merely mediated by the mixed $Ada$ Chern-Simons terms but may also occur through the current-dipole response ${\cal R}$ of composite fermions.

As derived in Appendix~\ref{app:B}, the free Dirac response functions without magnetic field have the following nonzero components:
\begin{align}
K^{\mu\nu} &= \begin{pmatrix} K^{00} & 0 \\ 0 & K^{xx} \end{pmatrix} \\
{\cal R}^{\nu i} &= \begin{pmatrix} 0 & {\cal R}^{02} \\ {\cal R}^{x1} & 0 \end{pmatrix} \\
\Delta^{ij} &= \begin{pmatrix} \Delta^{11} & 0 \\ 0 & \Delta^{22} \end{pmatrix} .
\end{align}
These response functions and their analytic continuation to real frequency that gives the retarded Dirac response are calculated in Appendix~\ref{app:B}. In terms of these components, we obtain the following results for the retarded response functions:
\begin{align}
\Pi^{0x}(\omega, q) &= \frac{-i q_y}{2 e \phi_0} , \label{eq:res0x} \\
\Pi^{00}(\omega, q) &= \biggl(\frac{q}{4\pi\hbar}\biggr)^2 \frac{Z_L(\omega, q)}{[K^{xx}(\omega, q)]^*} + q^2 \Delta^{11}(\omega, q) , \label{eq:density} \\
\Pi^{xx}(\omega, q) &= \biggl(\frac{q}{4\pi\hbar}\biggr)^2 \frac{Z_T(\omega, q)}{[K^{00}(\omega, q)]^*} + \omega^2 \Delta^{22}(\omega, q) , \label{eq:current}
\end{align}
with the dimensionless residue functions
\begin{align}
Z_L(\omega, q) &= |{\cal R}^{x1}(\omega, q))|^2 \label{eq:Zl} , \\
Z_T(\omega, q) &= |1 + \frac{\omega}{q} {\cal R}^{02}(\omega, q)|^2 .
\end{align}
Note that the absence of a term of order ${\cal O}(1)$ in the residue~\eqref{eq:Zl} is due to the cancellation of the Chern-Simons term with the tadpole correction as discussed following Eq.~\eqref{eq:Seff2}.

For reference, we also note the result for the density and transverse current response in the non-Galilean invariant theory [i.e., without the dipole term~\eqref{eq:dipole}] first derived by~\textcite{son15}. The density response is stated in Eq.~\eqref{eq:SDRPA1} and the transverse current response is
\begin{align}
\Pi_0^{xx}(\omega, q) &= \biggl(\frac{q}{4\pi\hbar}\biggr)^2 \frac{1}{[K^{00}(\omega, q)]^*} \label{eq:nonG2} .
\end{align}
There is no direct dipole response $\Delta$, and the residue terms $Z_L$ and $Z_T$ are equal to unity. As discussed in the Introduction, these response functions show a spurious behavior at finite frequencies, which is rectified when including the dipole correction. In the following sections, we will discuss the properties of the full response functions~\eqref{eq:res0x}-\eqref{eq:current} in detail.

Before we proceed, note that it is straightforward to extend our calculation to a more general case that includes an additional interaction potential~\cite{simon98,giuliani05}. On the level of the RPA, this term is taken into account as a Hartree correction that changes the effective vector potential by
\begin{align}
\Delta A_0(x) &= \int d^2x' \, V(x-x') \langle \delta j_0(x') \rangle .
\end{align}
Electrons are then assumed to respond to the Hartree potential in addition to the external vector potential $\delta A_\mu$. In Fourier space, we have $\langle \delta j \rangle = \Pi [\delta A + V \langle \delta j \rangle]$, where $\Pi^{\mu\nu}$ is the RPA response derived previously without the interaction potential and $V = \begin{psmallmatrix}
V({\bf q}) & 0 \\
0 & 0
\end{psmallmatrix}$ .
The full RPA response $\tilde{\Pi}$ is then linked to the response $\Pi$ by
\begin{align}
\tilde{\Pi}^{-1} &= \Pi^{-1} - V .
\end{align}
In particular, for the density response, we have $[\tilde{\Pi}^{00}]^{-1} = [\Pi^{00}]^{-1} - V({\bf q})$.

\subsection{Hall response function}

The first response~\eqref{eq:res0x} is the Hall response function, which is completely fixed by the $AdA$ Chern-Simons term and receives no corrections on the RPA level for the versions of the Son-Dirac theory discussed here. It is connected to the Hall conductivity by~\cite{giuliani05}
 \begin{align}
\sigma_H &= \lim_{\omega \to 0} \lim_{q_y \to 0} \frac{i e^2}{q_y} \Pi^{0x}(\omega, q_y) = \frac{e^2}{2 h} .
 \end{align}
 This result agrees with the exact long-wavelength limit predicted by particle-hole symmetry~\cite{kivelson97}. Particle-hole symmetry also predicts a subleading correction of order ${\cal O}(q^2)$, $\sigma_H = \frac{e^2}{2h} (1-q^2\ell_B^2/4)$ for $v_F q \ll \omega$~\cite{read11,levin17}, which is not reproduced by the RPA calculation. However, this problem is shared between HLR and the Son-Dirac theory in the form considered in this paper, and the latter theory incorporates this correction if half the action of a fully filled Landau level is added to the Son-Dirac action~\cite{levin17},
 \begin{align}
S &= - \frac{e}{\phi_0} \int_q A_0(-q) \frac{1 - q^{-q^2\ell_B^2/2} - q^2\ell_B^2/2}{q\ell_B} A_1(q) ,
 \end{align}
 with further modifications if an explicit Coulomb interaction is taken into account. Note that this term does not affect the density and current response function discussed in the next sections.

\subsection{Density response function}

Consider first the density response~\eqref{eq:SDRPA1} of the unmodified Son-Dirac theory, which is proportional to the inverse Dirac current response $K^{xx}$. In the long-wavelength limit, the Dirac current response takes the form (in dimensionless notation where $\nu = \omega\ell_B/v_F$, $x=q\ell_B$, and \mbox{$\hat{K}^{xx} = 2 \pi \hbar \ell_B K^{xx}/v_F$}):
\begin{align}
\hat{K}^{xx}(\nu, {\bf x}) &= - \dfrac{1}{2} + \dfrac{\nu}{8} \ln \biggl| \dfrac{2+\nu}{2-\nu} \biggr|  + i \dfrac{\pi \nu}{8} \Theta(\nu-2) . \label{eq:Kxxq0}
\end{align}
The full expression valid at all momenta is given in Appendix~\ref{app:B}. The real part of this response  crosses zero at a finite frequency $\nu = 1.66711$, which, as discussed in the Introduction, gives rise to a spurious collective mode in the dynamic structure factor of the non-Galilean invariant theory (cf. Fig.~\ref{fig:1}). In addition, there is incoherent spectral weight corresponding to interband transitions with asymptotic weight ${\cal O}(x^2/\nu)$ that leads to a UV-divergence of the $f$-sum rule~\eqref{eq:fsumrule}. In the static limit, the Dirac current response reads
\begin{align}
\hat{K}^{xx}(\nu, {\bf x}) &= \biggl( - \frac{\pi x}{8} - \frac{\sqrt{x^2-4}}{2x} + \frac{x}{4} \arcsin \tfrac{2}{x}\biggr) \Theta(x-2) .
\end{align}
This function vanishes for momenta $q\ell_B < 2$, which is linked to a divergent orbital susceptibility of Dirac electrons at finite detuning~\cite{principi09}. For the Son-Dirac theory, this implies a divergent static response function, which is changed to a linear function of momentum if a Coulomb interaction is included, the latter result being consistent with HLR theory~\cite{hlr93}. Finally, in the Fermi-liquid scaling regime at small momentum and frequency, we have (defining the scaling variable $s=\omega/v_Fq$)
\begin{align}
\lim_{\omega,q \to 0} S^{{\rm SD}}(\omega, q) &= \frac{q^2 \ell_B}{2\pi^2 \hbar v_F} \frac{\sqrt{1-s^2}}{4 s} \Theta(s<1) . \label{eq:Sqindirect}
\end{align}
Indeed, neglecting the high-frequency response present in the Son-Dirac theory, the density response~\eqref{eq:Sqindirect} at small frequencies is equal to the HLR result (Appendix~\ref{app:hlr}) if we identify the Fermi velocity in the Son-Dirac theory and the effective mass parameter in HLR theory in the natural way, $v_F = \hbar k_F/m^*$. The real parts of the density response are equal as well, such that both theories predict a (sub)diffusive mode with frequency $\nu = - i x^2 (\frac{\alpha}{2} + \frac{2 x}{3})$~\cite{hlr93,son15}, where $\alpha = e^2/\hbar v_F$ is the dimensionless Coulomb interaction strength~\cite{hofmann14}. 
  Moreover, evaluating the $f$-sum rule using the low-frequency response~\eqref{eq:Sqindirect}, we obtain the scaling
\begin{align}
\bar{f}(q) &= \frac{\pi}{4} (q\ell_B)^4 , \label{eq:fsum}
\end{align}
as required in the LLL from the Girvin-Mac Donald-Platzman algebra~\cite{girvin86}. Evaluating the static structure factor in this limit gives an asymptotic IR-divergent scaling $S(q) = {\cal O}((q\ell_B)^3 \ln (q\ell_B))$, consistent with a compressible phase at half-filling~\cite{girvin86} and again the same as for HLR theory~\cite{hlr93,murthy98} (cf. Appendix~\ref{app:hlr}). Of course, these calculations ignore the spurious interband excitations and collective modes discussed previously, which  give a divergent contribution to the sum rules.

\begin{figure}[t]
\scalebox{0.7}{\includegraphics{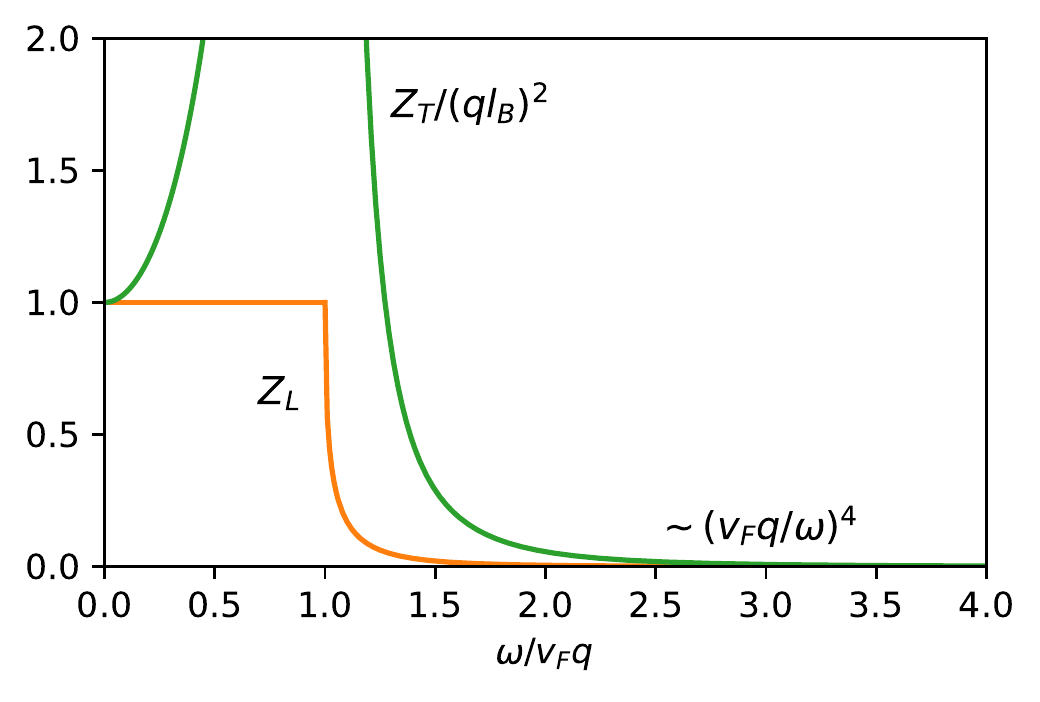}}
\caption{Small-momentum and small-energy scaling limit of the dipole residue terms $Z_L$ and $Z_T$ as a function of the scaling parameter $s=\omega/v_Fq$. Both residues decay rapidly for frequencies larger than $\omega > v_Fq$, which leads to a strong suppression of the response at large frequencies $\omega \sim \hbar v_F/\ell_B$.}
\label{fig:residue}
\end{figure}

\begin{figure}[t]
\scalebox{0.88}{\includegraphics{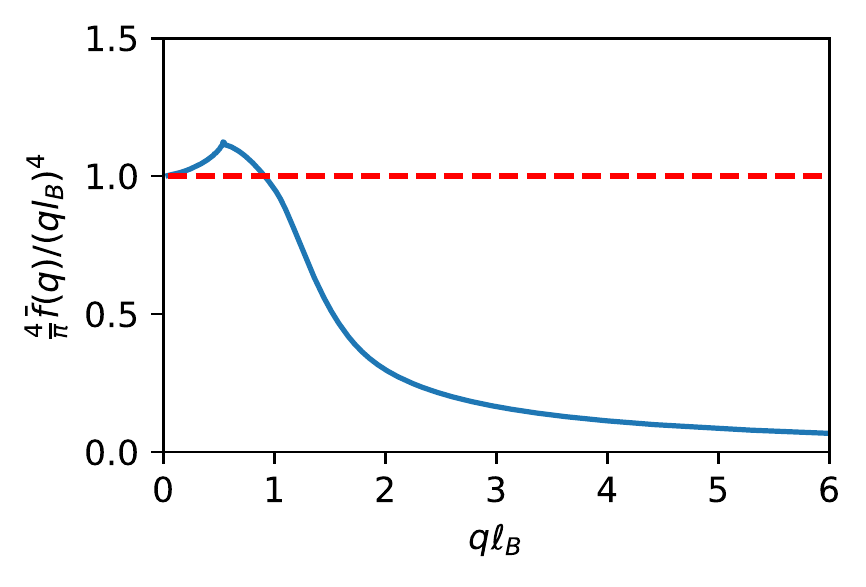}}
\caption{$f$-sum rule computed using the indirect contribution to the density response of the modified Son-Dirac theory. The dashed line indicates the limit~\eqref{eq:fsum}.}
\label{fig:sumrules}
\end{figure}

Consider now the density response~\eqref{eq:density} of the modified Son-Dirac theory, which involves the density response of the original Son-Dirac theory modified by a dipole residue factor $Z_L$ as well as a direct response term that describes the direct coupling of dipoles to the external probe. Crucially, the dipole residue $Z_L$ [Eq.~\eqref{eq:Zl}] leaves the low-frequency response of the Son-Dirac theory discussed above unchanged but strongly suppresses the spurious large-frequency response. To see this, consider the low-momentum scaling form of~\eqref{eq:Zl}:
\begin{align}
&\lim_{\omega,q \to 0}  Z_L(\omega, q) \nonumber \\*
&\ = \begin{cases}
1 , & s<1 , \\[2ex]
1 - 8 s^2 + 8 s^4 - 4 s (2 s^2-1) \sqrt{s^2-1} , & s>1 .
\end{cases} \label{eq:Zllimit}
\end{align}
This result is shown as a continuous orange line in Fig.~\ref{fig:residue}, and the full result for this part of the dynamic structure factor is shown in Fig.~\ref{fig:2}. For $\omega < v_F q$, the residue term in Eq.~\eqref{eq:Zllimit} is unity [such that the dynamic structure factor in the scaling regime is unchanged from the Son-Dirac results~\eqref{eq:Sqindirect}] and then decays very quickly [on a scale of ${\cal O}(q)$] to zero with an asymptotic form $(1/2s)^4 = v_F^4 q^4/16\omega^4$. The spurious interband transitions and the collective mode (which set in at a much larger frequency) are then suppressed by four further orders of magnitude as ${\cal O}(q^6)$ at small momentum. Indeed, as is apparent from Fig.~\ref{fig:2}, this suppression of spectral weight at $\omega > v_F q$ holds for all momenta. In particular, if one restricts the response to the Chern-Simons contribution, the $f$-sum rule is finite for all momenta, which is shown in Fig.~\ref{fig:sumrules}. The $f$-sum rule takes the form (52) at small momentum, has a kink at the point where the spurious collective mode joins the particle-hole continuum, and then crosses over to an asymptotic power-law scaling ${\cal O}((q\ell_B)^4)$. 

\begin{figure}[t]
\scalebox{0.845}{\includegraphics{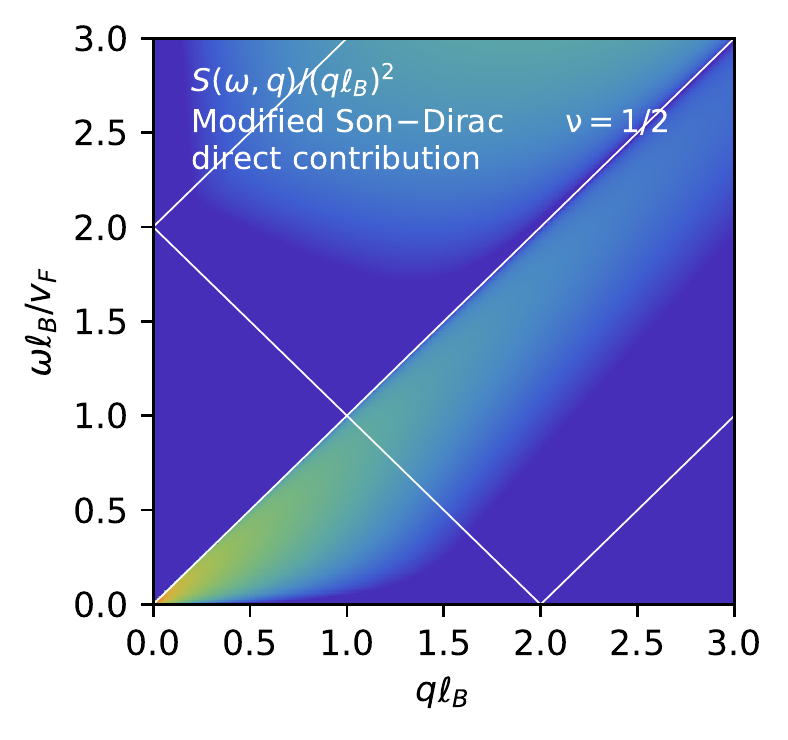}}
\caption{Spectral function of the direct dipole response $q^2 \Delta^{11}$ as a function of momentum and frequency.}
\label{fig:dipoledensity}
\end{figure}

The full RPA response~\eqref{eq:density} contains an additional direct dipole-dipole response term $q^2 \Delta^{11}$. In the small-momentum limit, this response reads (in dimensionless form \mbox{$\hat{\Delta}^{11} = 2 \pi \hbar \Delta^{11}/v_F \ell_B$})
\begin{align}
&\hat{\Delta}^{11}(\nu, {\bf x}) \nonumber \\*
&  = - \dfrac{x^4 (4 + 3 \nu^2)}{8 \nu^2} + \dfrac{3 x^4 \nu}{32} \ln \biggl| \dfrac{2+\nu}{2-\nu} \biggr| + i \dfrac{3 i \pi \nu x^4}{32} \Theta(\nu-2)  \label{eq:K11q0}
\end{align}
with the full result for all momenta and frequencies computed and stated in Appendix~\ref{app:B}. The  corresponding contribution to the dynamic structure factor is shown in Fig.~\ref{fig:dipoledensity}. Interband transition are suppressed as ${\cal O}(q^4)$ in the low-momentum limit and thus sub-leading compared to the original Son-Dirac theory, where they are of order ${\cal O}(q^2)$. However, beyond this order, due to the linear frequency behavior~\eqref{eq:K11q0}, sum rules at arbitrary momentum are no longer finite, and higher-order corrections to the Son-Dirac theory will have to be taken into account to obtain a consistent theory of the half-filled Landau level. At quadratic order, where the high-frequency response is absent, the direct dipole response will affect the Fermi liquid scaling regime, where it contributes a term
\begin{align}
\lim_{\omega,q \to 0} S^{{\rm MSD},d}(\omega, q) &= \frac{q^2 \ell_B}{2\pi^2 \hbar v_F} 4 s \sqrt{1-s^2} \, \Theta(s<1) 
\end{align}
to the dynamic structure factor in addition to Eq.~\eqref{eq:Sqindirect}. The static response of the full result is unchanged, but there is added spectral weight near the particle-hole threshold $\omega \simeq v_F q$. Note that this direct dipole contribution to the RPA response of the modified Son-Dirac theory is not contained in the RPA response of the modified HLR theory.

\begin{figure}[t]
\scalebox{0.845}{\includegraphics{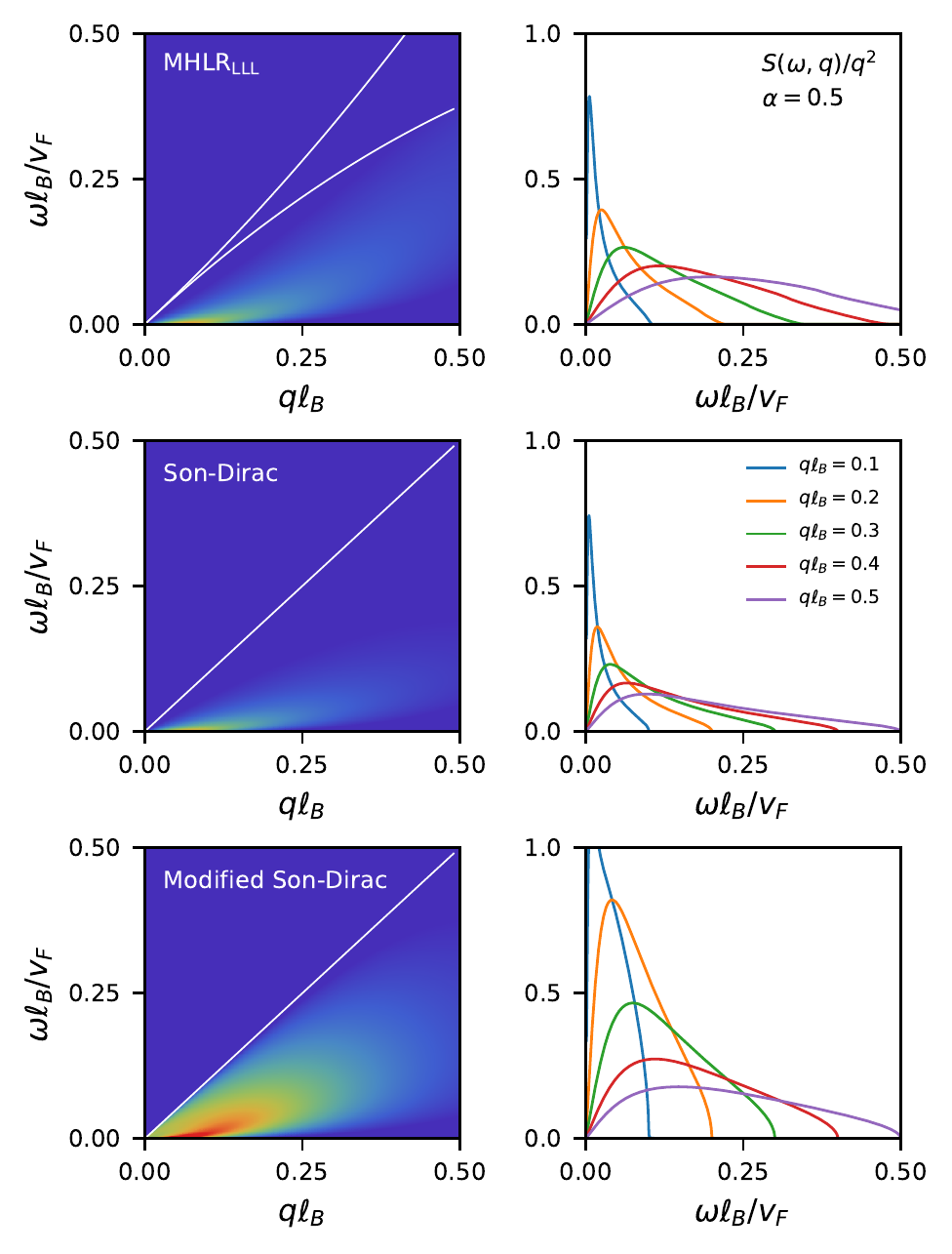}}
\caption{Dynamic structure factor of the modified HLR theory (top row), the Son-Dirac theory (middle row) and the modified Son-Dirac theory (bottom row) for a fixed Coulomb interaction strength $\alpha = 0.5$. While HLR and Son-Dirac theories make similar predictions, there is a difference compared to the modified Son-Dirac theory, which is marked by an enhanced incoherent spectral weight near the particle-hole threshold.}
\label{fig:scaling}
\end{figure}

To illustrate the difference between the RPA response of the Son-Dirac theory, the modified Son-Dirac theory, as well as the modified HLR theory, we show in Fig.~\ref{fig:scaling} the dynamic structure factor of the modified HLR theory, the Son-Dirac theory, and the modified Son-Dirac theory (top to bottom row) including a dimensionless Coulomb interaction strength $\alpha = 0.5$ (note that the HLR interaction parameter is identified as $\hat{r}_s = 2 \alpha$ when setting $v_F = \hbar k_F/m^*$). The panels on the left-hand side show a density plot of the dynamic structure factor, and plots on the right-hand side show it as a function of frequency for five fixed momenta $q\ell_B = 0.1, 0.2, 0.3, 0.4,$ and $0.5$. For small momenta, the Coulomb interaction is subleading and the dominant feature is the low-frequency divergence~\eqref{eq:Sqindirect} that is also sketched at the bottom plot of Fig.~\ref{fig:1}. This divergence is cut off at finite momenta by the Coulomb interaction, such that the response is linear at small frequencies. At larger frequency, there is additional incoherent spectral weight that slowly decays up to the phase-space boundary $\omega = v_F q$. As is apparent from the figure and discussed in this section, the RPA results for the Son-Dirac theory and the modified HLR theory in this regime are very similar. There is, however, a clearly visible difference compared to the modified Son-Dirac theory, which has increased incoherent weight near threshold. This distinguishes the RPA response of the modified Son-Dirac theory from the RPA response of the modified HLR theory.

\subsection{Transverse current response function}

\begin{figure*}[t]
\scalebox{0.7}{\includegraphics{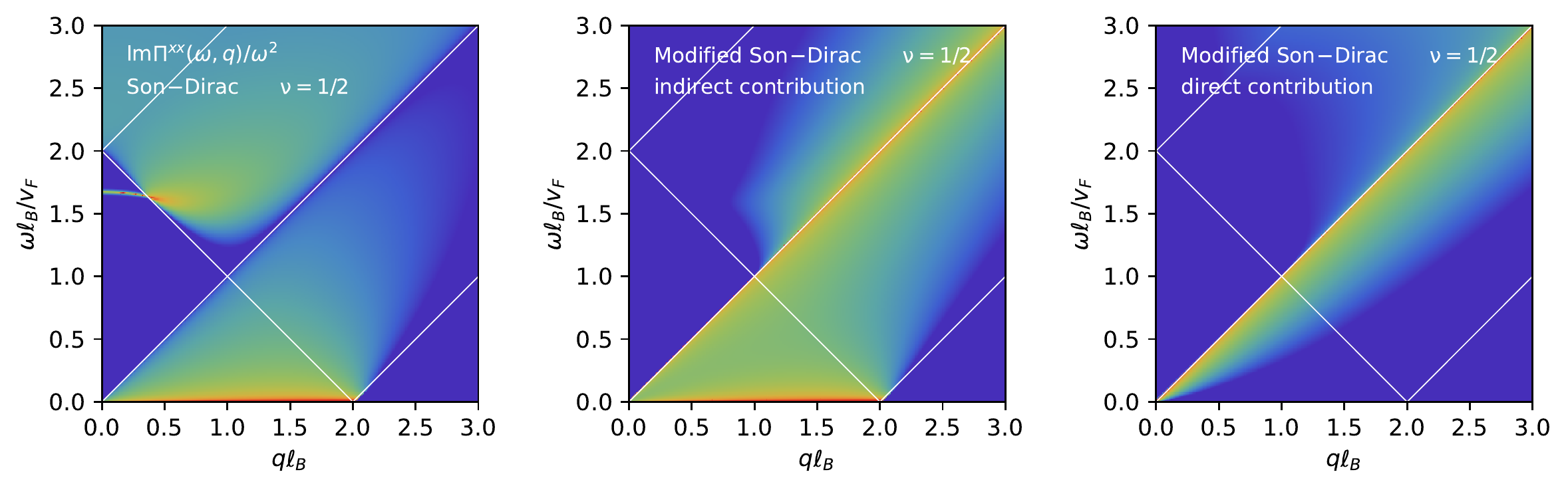}}
\caption{Spectral function ${\rm Im} \, \Pi^{xx}$ of the transverse current response as a function of momentum and frequency for (a) the non-Galilean invariant theory [Eq.~\eqref{eq:nonG2}]; (b) the full indirect response [Eq.~\eqref{eq:current}] including the dipole residue; and (c) the direct dipole response function $\omega^2 \Delta^{22}$.}
\label{fig:transversecurrent}
\end{figure*}

A similar discussion to that for the density response function applies to the transverse current response function~\eqref{eq:current}. Figure~\ref{fig:transversecurrent} shows (from left to right) the spectral function ${\rm Im} \, \Pi^{xx}$ of the non-Galilean invariant response~\eqref{eq:nonG2}, the indirect Galilean-invariant response that arises from the coupling to the Chern-Simons field [first term in Eq.~\eqref{eq:current}], as well as the direct dipole response [second term in Eq.~\eqref{eq:current}]. In the small-momentum limit, the Dirac density response is given by (in dimensionless form \mbox{$\hat{K}^{00} = 2 \pi \hbar v_F \ell_B K^{00}$})
\begin{align}
\hat{K}^{00}(\nu, {\bf x}) &= - \dfrac{x^2}{2 \nu^2} + \dfrac{x^2}{8 \nu} \ln \biggl| \dfrac{2+\nu}{2-\nu} \biggr| + \dfrac{i \pi x^2}{8 \nu}  \Theta(\nu-2) , \label{eq:K00q0}
\end{align}
which, as for the density response, gives rise to a pole at $\nu = 1.66711$ with contributions from interband transitions diverging linearly at large frequency as ${\cal O}(\omega q^0)$. The small-momentum limit of the dipole residue $Z_T$ is given by
\begin{align}
&\lim_{\omega, q \to 0}  Z_T(\omega, q) &= \begin{cases}
\dfrac{1 + 3 s^2}{1 - s^2} , & s\leq 1 , \\[2ex]
\Bigl(1 + 2s^2 - \frac{2s^3}{\sqrt{s^2-1}}\Bigr)^2 , & s>1 ,
\end{cases}  \label{eq:Ztlimit}
\end{align}
which at large frequencies decays as $9 q^4/16\omega^4$, thus suppressing the collective mode and interband contributions. This result is shown as a continuous green line in Fig.~\ref{fig:residue}. Different from the density response function, the dipole residue is not equal to unity in the Fermi-liquid scaling region and thus the response differs from the non-Galilean result~\eqref{eq:nonG2}. Indeed, the corresponding spectral function takes the form
\begin{align}
\lim_{\omega,q \to 0} {\rm Im} \, \Pi^{xx,a}(\omega, q) &= \frac{v_F q^2 \ell_B}{2\pi \hbar } \frac{s (1+3s^2)}{4 \sqrt{1 - s^2}} \, \Theta(s<1) .
\end{align}
The small-momentum limit of the additional direct dipole response $\omega^2 \Delta^{22}$ is  (in dimensionless form \mbox{$\hat{\Delta}^{22} = 2 \pi \hbar \Delta^{22}/v_F \ell_B$})
\begin{align}
&\hat{\Delta}^{22}(\nu, {\bf x}) \nonumber \\*
&= - \dfrac{x^2 (12 + \nu^2)}{8} + \dfrac{x^2 \nu^3}{32} \ln \biggl| \dfrac{2+\nu}{2-\nu} \biggr| + i \dfrac{i \pi \nu^3 x^2}{32} \Theta(\nu-2) \label{eq:K22q0}
\end{align}
with the full momentum and frequency dependence presented in Appendix~\ref{app:B}. In the low-frequency and momentum scaling limit, the dipole response contributes a term
\begin{align}
\lim_{\omega,q \to 0} {\rm Im} \, \Pi^{xx,b}(\omega, q) &= \frac{v_F q^2 \ell_B}{2\pi \hbar } \dfrac{4 s^3}{\sqrt{1-s^2}} \, \Theta(s<1) 
\end{align}
to the spectral function. As for the density response, this contribution is negligible in the static limit but changes the response near threshold. The enhancement of spectral weight is due to both the dipole residue, which diverges as $Z_T \sim 2/|1-s|$ near $s \simeq 1$, as well as the direct dipole response $\Delta ^{22}$.

\section{Summary}\label{eq:summary}

In summary, we have discussed the response of the Son-Dirac theory of the half-filled Landau level using the random phase approximation. If a dipole correction is included that renders the Son-Dirac theory Galilean invariant, we find that the response is free of spurious high-frequency excitations, which are natural features in the response of Dirac materials but not expected for a theory of the LLL. Furthermore, while the response of the Son-Dirac theory reproduces many features of HLR theory at small frequencies and momenta, the dipole term increases the response near the particle-hole boundary. This is a prediction of the Son-Dirac theory within the RPA that differs from the modified HLR theory within the RPA. In future work, it would be interesting to extend the present calculation to other gapless filling fractions $\nu=1/2n$ and $\nu=(2n-1)/2n$, for which a generalization of the Son-Dirac theory has been proposed~\cite{goldman18,wang19,nguyen21}, and to states in the Jain sequence $\nu = n/(2n+1)$ and $\nu=(n+1)/(2n+1)$, which are described by fully filled Landau levels of composite fermions~\cite{son15}. 

\begin{acknowledgements}
I thank G. M\"oller and C. Turner for discussions, and N. Cooper for discussions and comments on the manuscript. 
This work is supported by Vetenskapsr\aa det (grant number 2020-04239).
\end{acknowledgements}

\appendix

\section{Halperin-Lee-Read response in the LLL}\label{app:hlr}

This appendix collects for reference results for the electron response within HLR theory at half-filling restricted to the lowest Landau level, which are compared in the main text with the findings of this paper.

The starting point is a modified version of HLR theory that contains as a parameter an interaction-renormalized mass $m^*$ instead of the bare mass $m$ (for a review, see Ref.~\cite{simon98}). In order to restore Galilean invariance, this theory must include a Fermi-liquid back-flow term, the strength of which is set by the $p$-wave Fermi liquid parameter $F_1 = \frac{m}{m^*} - 1$. Within this framework, response functions can be computed in closed analytic form on the level of the RPA and linked to the response functions of the nonrelativistic two-dimensional electron gas (2DEG). The LLL limit corresponds to the limit of fixed filling fraction $\nu \sim {\cal O}(m^0)$ and diverging cyclotron frequency $\omega_c \sim {\cal O}(m^{-1})$. This is the limit of vanishing bare mass, $m\to 0$ and $F_1 \to \infty$, respectively. Units are now set by the renormalized mass, such that energies are measured in units of $\hbar \tilde{\omega}_c = \hbar e B/m^*$ with a dimensionless frequency $\nu = \hbar \omega/\hbar \tilde{\omega}_c$ and wave number $x = q \ell_B$ (note that this unit of energy differs from the standard definition of a 2DEG $E_F = \hbar^2k_F^2/2m^*$ with Fermi momentum $k_F=1/\ell_B$ by a factor of $2$). 

\begin{figure}[t]
\begin{tabular}{cc}\raisebox{-1cm}{
\multirow{2}{*}[27mm]{
\hspace{-0.3cm}\scalebox{0.5}{\raisebox{2cm}{\includegraphics{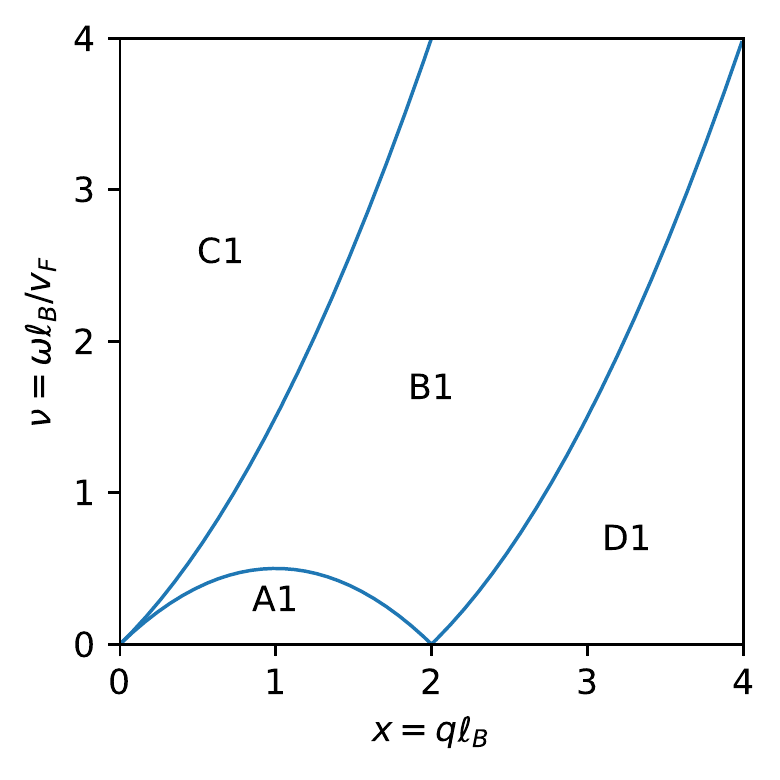}}}}}\vspace{-1cm}
&
\scalebox{0.48}{\includegraphics{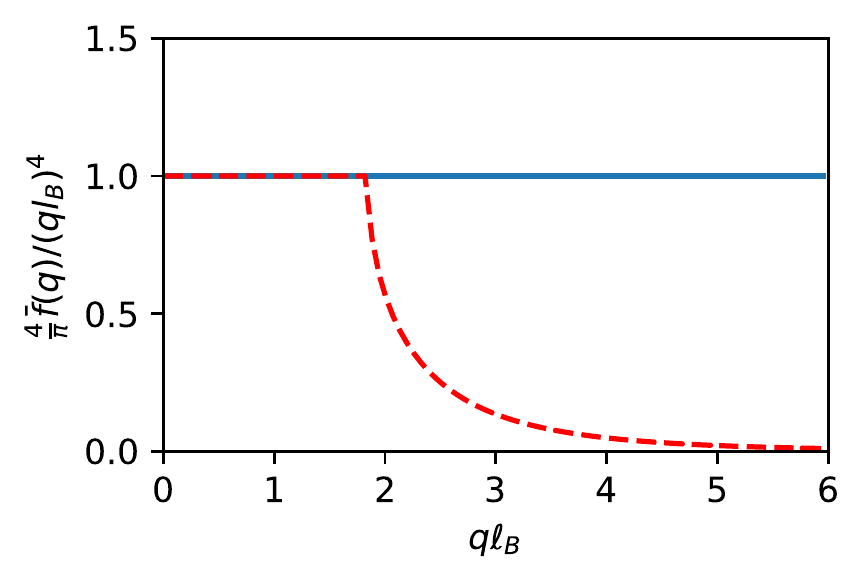}} \\
&
\hspace{0.2cm}\scalebox{0.47}{\includegraphics{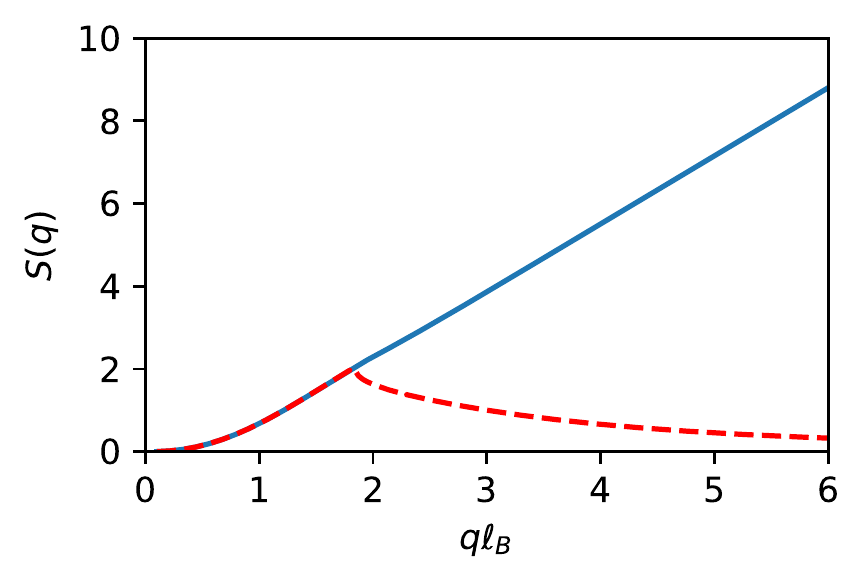}} 
\vspace{-0.2cm}
\end{tabular}
\caption{Left panel: Phase-space regions used to parametrize the free-particle response functions of 2D electrons. Right panels: $f$-sum rule (top) and static structure factor (bottom) of HLR theory projected onto the lowest Landau level. The continuous blue lines indicate the full result, and the dashed red lines exclude a collective mode that decouples from the continuum response at finite momentum.}
\label{fig:HLR}
\end{figure}

At half-filling, the dimensionless density response is~\cite{simon98}
\begin{align}
&[\hat{\Pi}^{00}]^{-1}(\nu, {\bf x}) = \frac{m^*}{2\pi \hbar^2} [{\Pi}^{00}]^{-1} \nonumber \\*
&=  [\hat{K}_{00}^0]^{-1} - \frac{2 \hat{r}_s}{x} - \frac{2 \nu^2}{x^2} - \frac{4}{x^2} \Bigl([\hat{K}^0_{xx}]^{-1} - 2\Bigr)^{-1} , \label{eq:dimlessdensityhalf}
\end{align}
where $\hat{r}_s = r_0/\tilde{a}_0$ is the dimensionless Coulomb interaction strength, defined as the ratio of average electron spacing $r_0=1/\sqrt{\pi n}$ and interacting Bohr radius $\hat{a}_0 = \hbar^2/m^*e^2$. The expression~\eqref{eq:dimlessdensityhalf} contains the standard 2DEG response functions~\cite{giuliani05}
\begin{align}
&{\rm Re} \hat{K}^{\alpha\alpha}(\nu, {\bf x}) = - h^\alpha(\nu,x) \nonumber \\*
& - \frac{1}{x} \begin{cases}
0 & {\rm A1} , \\[2ex]
g^\alpha_>\bigl(\dfrac{\nu}{x} + \dfrac{x}{2}\bigr) & {\rm B1} , \\[2ex]
g^\alpha_>\bigl(\dfrac{\nu}{x} + \dfrac{x}{2}\bigr) - g^\alpha_>\bigl(\dfrac{\nu}{x} - \dfrac{x}{2}\bigr) & {\rm C1} , \\[2ex]
g^\alpha_>\bigl(\dfrac{\nu}{x} + \dfrac{x}{2}\bigr) + g^\alpha_>\bigl(\dfrac{\nu}{x} - \dfrac{x}{2}\bigr) & {\rm D1} ,
\end{cases}
\end{align}
\begin{align}
{\rm Im} \hat{K}^{\alpha\alpha}(\nu, {\bf x}) &= - \frac{1}{x}
\begin{cases}
g^\alpha_<\bigl(\dfrac{\nu}{x} + \dfrac{x}{2}\bigr) - g^\alpha_<\bigl(\dfrac{\nu}{x} - \dfrac{x}{2}\bigr) & {\rm A1} , \\[2ex]
- g^\alpha_<\bigl(\dfrac{\nu}{x} - \dfrac{x}{2}\bigr) & {\rm B1} , \\[2ex]
0 & {\rm C1} , \\[2ex]
0 & {\rm D1} ,
\end{cases}
\end{align}
where the four regions are sketched in Fig.~\ref{fig:HLR}(a), with
\begin{align}
h^\alpha(\nu,x) &= \begin{cases}
-1 , & \alpha = 0 , \\[2ex]
\dfrac{x^2}{12} + \dfrac{\nu^2}{x^2} , & \alpha = x ,
\end{cases} \\
g^\alpha_>(\nu,x) &= \begin{cases}
\sqrt{z^2-1} , & \alpha = 0 , \\[2ex]
\dfrac{1-z^2}{3} \sqrt{z^2-1} , & \alpha = x ,
\end{cases} \\
g^\alpha_<(\nu,x) &= \begin{cases}
\sqrt{1-z^2} , & \alpha = 0 , \\[2ex]
\dfrac{1-z^2}{3} \sqrt{1-z^2} , & \alpha = x .
\end{cases} 
\end{align}
The full results is shown in the inset of Fig.~\ref{fig:1}. It is interesting to note that there is a collective mode that emerges at larger momentum $q\ell_B \gtrsim 1.9$.

The small-momentum behavior of the full HLR response is (without a Coulomb term, i.e., $\hat{r}_s = 0$)
\begin{align}
\hat{\Pi}^{00}(\nu, {\bf x}) &= 
\begin{cases}
\biggl(- \dfrac{1}{4} + i \dfrac{\sqrt{1-s^2}}{4s}\biggr) x^2 , & s<1 , \\[2ex]
\biggl(- \dfrac{1}{4} + \dfrac{\sqrt{s^2-1}}{4s}\biggr) x^2 , & s>1 .
\end{cases} \label{eq:PiHLRexp}
\end{align}
The results for the $f$-sum rule and the static structure factor within HLR theory are shown in Figs.~\ref{fig:HLR}(b) and~\ref{fig:HLR}(c), where the blue continuous lines denote the full contribution of both the continuum and the collective mode that emerges at larger momenta (cf. the inset in Fig.~\ref{fig:1}), and the red dashed line excludes the collective mode. The projected $f$-sum rule takes the value~\eqref{eq:fsum} for all momenta. Moreover, the static structure factor vanishes as $S(x) = {\cal O}(x^3 \ln x)$, the same as discussed in the main text after Eq.~\eqref{eq:fsum}.

\section{Linear response of composite Dirac fermions}\label{app:B}

In the main text, the response of electrons in the half-filled lowest Landau level is expressed via the RPA in terms of six linearly independent response functions of noninteracting 2D Dirac fermions. In this appendix, we compute these response functions in closed analytical form. Two of these Dirac response functions --- the density and current response --- are discussed in the graphene literature as well~\cite{hwang07,wunsch06,principi09}, while to the best of our knowledge the four independent response functions involving the dipole term are new. Here, we derive all response functions for completeness.

Evaluating the frequency integral in Eqs.~\eqref{eq:defK}-\eqref{eq:defDelta} gives the standard Lindhard form of the response functions,
\begin{align}
&M^{\mu\nu}(i\omega,{\bf q}) \nonumber \\*
&= 
\int \frac{d{\bf k}}{(2\pi)^2} \sum_{ss'} \, \frac{f(E_{s'}({\bf k}')) - f(E_s({\bf k}))}{E_{s'}({\bf k}') - E_s({\bf k}) - i \hbar \omega} \, F_{ss'}^{\mu\nu}({\bf k}, {\bf k}') ,
\label{eq:poldef}
\end{align}
where $M$ is either one of the response functions $K$, ${\cal R}$, and $\Delta$ discussed in the main text, $f(E_s({\bf k})) = n_F(E_s({\bf k}) - \mu)$ is the Fermi-Dirac distribution for a system with chemical potential $\mu$ ($\mu = E_F = \hbar v_F/\ell_B$ at zero temperature), $E_s({\bf k}) = \pm \hbar v_F k$ is the single-particle energy of a Dirac state with wave vector ${\bf k}$ and chiral band index $s = \pm 1$, and $F_{ss'}^{\mu\nu}({\bf k}, {\bf k}')$ is the matrix element of the vertex terms between two single-particle eigenstates $|{\bf k} s \rangle$ and $|{\bf k}' s'\rangle$. Here, $F_{ss'}^{\mu\nu}({\bf k}, {\bf k}')$ is formed by an appropriate product of two of the following single-particle matrix elements:
\begin{align}
\langle {\bf k}' s'| I |  {\bf k} s\rangle &= \frac{1}{2} (1 + s s' e^{i (\phi - \phi')}) , \label{eq:mat1} \\
\langle {\bf k}' s'| \sigma^x | {\bf k} s\rangle &= \frac{1}{2} (s e^{i \phi} + s' e^{- i \phi'}) , \\
\langle {\bf k}' s'| \tfrac{-i \overset\leftrightarrow{\partial_x}}{2} |  {\bf k} s\rangle &= \frac{k \cos \phi + k' \cos \phi'}{2} (1 + s s' e^{i (\phi - \phi')}) , \\
\langle {\bf k}' s'| \tfrac{-i \overset\leftrightarrow{\partial_y}}{2} |  {\bf k} s\rangle &= \frac{k \sin \phi + k' \sin \phi'}{2} (1 + s s' e^{i (\phi - \phi')}) ,
\end{align}
where $\phi$ and $\phi'$ are the angles of the vectors ${\bf k}$ and ${\bf k}'$ in the complex plane.

It is convenient to split the response into an intrinsic part $M_{-}^{\mu\nu}$ that describes the response of a system at $\mu = 0$ (i.e., where the Fermi level is precisely at the Dirac point) as well as an extrinsic part $M_{+}^{\mu\nu}$ that contains the correction of finite detuning:
\begin{widetext}
\begin{align}
M^{\mu\nu}(i \omega, {\bf q}) &= M_{-}^{\mu\nu}(\omega, {\bf q}) + M_{+}^{\mu\nu}(\omega, {\bf q}) , \\*
M_{-}^{\mu\nu}(i \omega, {\bf q}) &= \hspace{-0.1cm}\int \hspace{-0.2cm} \frac{d{\bf k}}{(2\pi)^2} \biggl[
\frac{f(E_{-}({\bf k}')) - f(E_{-}({\bf k}))}{E_{-}({\bf k}') - E_{-}({\bf k}) - i \hbar \omega} F^{\mu\nu}_{--}({\bf k},{\bf k}') 
+ \frac{f(E_{-}({\bf k}')) F^{\mu\nu}_{+-}({\bf k},{\bf k}') }{E_{-}({\bf k}') - E_{+}({\bf k}) - i \hbar \omega}
- \frac{f(E_{-}({\bf k})) F^{\mu\nu}_{-+}({\bf k},{\bf k}')}{E_{+}({\bf k}') - E_{-}({\bf k}) - i \hbar \omega} \biggr] \label{eq:Km} , \\*
M_{+}^{\mu\nu}(i \omega, {\bf q}) &=\hspace{-0.1cm}\int \hspace{-0.2cm} \frac{d{\bf k}}{(2\pi)^2} \biggl[
\frac{f(E_{+}({\bf k}')) - f(E_{+}({\bf k}))}{E_{+}({\bf k}') - E_{+}({\bf k}) - i \hbar \omega} F^{\mu\nu}_{++}({\bf k},{\bf k}') 
+ \frac{f(E_{+}({\bf k}'))  F^{\mu\nu}_{-+}({\bf k},{\bf k}') }{E_{+}({\bf k}') - E_{-}({\bf k}) - i \hbar \omega}
- \frac{f(E_{+}({\bf k}))  F^{\mu\nu}_{+-}({\bf k},{\bf k}')}{E_{-}({\bf k}') - E_{+}({\bf k}) - i \hbar \omega}\biggr] . \label{eq:Kp}
\end{align}
\end{widetext}
In the following, we evaluate both contributions in turn, computing first the Euclidean response at imaginary frequency $M^{\mu\nu}(i \omega, {\bf q})$ and then performing the analytic continuation $i\omega \to \omega + i0$ to obtain the retarded response. We shall use dimensionless variables 
\begin{align}
{\bf q} &= k_F {\bf x} =  {\bf x}/\ell_B , \\
\hbar \omega &= E_F \nu = \hbar v_F \nu/\ell_B .
\end{align}
In addition, we introduce dimensionless response functions, which we shall indicate by a hat, as follows:
\begin{align}
{K}^{00}(i \omega, {\bf q}) &= {\cal N}_0 \hat{K}^{00}(\nu = \tfrac{ \omega}{v_F k_F}, {\bf x} = \tfrac{q}{k_F}) , \\
{K}^{xx}(i \omega, {\bf q}) &= v_F^2 {\cal N}_0 \hat{K}^{xx}(\nu = \tfrac{ \omega}{v_F k_F}, {\bf x} = \tfrac{q}{k_F}) , \\
{\Delta}^{11}(i \omega, {\bf q}) &= \ell_B^2 {\cal N}_0 \hat{\Delta}^{11}(\nu = \tfrac{ \omega}{v_F k_F}, {\bf x} = \tfrac{q}{k_F}) , \\
{\Delta}^{22}(i \omega, {\bf q}) &= \ell_B^2 {\cal N}_0 \hat{\Delta}^{22}(\nu = \tfrac{\omega}{v_F k_F}, {\bf x} = \tfrac{q}{k_F}) , \\ 
{\cal R}^{02}(i \omega, {\bf q}) &= v_F^{-1} \hat{\cal R}^{02}(\nu = \tfrac{ \omega}{v_F k_F}, {\bf x} = \tfrac{q}{k_F}) , \\
{\cal R}^{x1}(i \omega, {\bf q}) &= \hat{\cal R}^{x1}(\nu = \tfrac{ \omega}{v_F k_F}, {\bf x} = \tfrac{q}{k_F}) ,
\end{align}
where ${\cal N}_0$ is the density of states of a noninteracting system at the Fermi surface
\begin{align}
{\cal N}_0 &= \frac{k_F}{2\pi \hbar v_F} =  \frac{1}{2\pi \hbar v_F \ell_B}.
\end{align}
Note that if we define an effective mass $m$ by $v_F = \hbar k_F/m$, this is equivalent to the noninteracting density of states of the 2DEG, for which ${\cal N}_0^{\rm 2DEG} = m/2\pi\hbar^2$.

\subsection{Intrinsic response}

In this section, we consider the intrinsic contribution to the response $M_{-}^{\mu\nu}$, which is the full response if the Fermi energy is at the Dirac point. In this case, only interband transitions (which have \mbox{$ss'=-1$}) contribute, Eq.~\eqref{eq:Km}, such that
\begin{align}
M_{-}^{\mu\nu}(i \omega, {\bf q}) &= - \int \frac{d{\bf k}}{(2\pi)^2} \biggl[
\frac{F^{\mu\nu}_{+-}({\bf k},{\bf k}')}{E_{-}({\bf k}') - E_{+}({\bf k}) - i \hbar \omega}  \nonumber \\*
&\qquad
- \frac{F^{\mu\nu}_{-+}({\bf k},{\bf k}')}{E_{+}({\bf k}') - E_{-}({\bf k}) - i \hbar \omega} \biggr] . \label{eq:intrinsic}
\end{align}
To evaluate this part, it is convenient to shift the integration variable ${\bf k} \to {\bf k} - {\bf q}/2$ with ${\bf q} = (0,q)$ and transform to an elliptic coordinate system with $\pm {\bf q}/2$ at the focus points~\cite{throckmorton15},
\begin{align}
k_x &= \frac{q}{2} \sinh \mu \sin \nu \\
k_y &= \frac{q}{2} \cosh \mu \cos \nu ,
\end{align}
where $\mu>0$ and $-\pi<\nu<\pi$. The Jacobian of the transformation is
\begin{align}
\biggl|\frac{\partial(k_x,k_y)}{\partial(\mu,\nu)}\biggr| &
= \frac{q^2}{4} (\sinh^2 \mu + \sin^2 \nu) .
\end{align}
In these coordinates, the matrix elements are
\begin{align}
& \langle {\bf k} + \tfrac{{\bf q}}{2} s'| I |  {\bf k} - \tfrac{{\bf q}}{2} s\rangle \nonumber \\
&= \frac{1}{\sinh \mu - i \sin \nu} \times \begin{cases} \sinh \mu , & s s' = + 1 , \\[1ex] - i \sin \nu , & s s' = - 1 ,\end{cases} \label{eq:mat1} \\
& \langle {\bf k} + \tfrac{{\bf q}}{2} s'| \sigma^x | {\bf k} - \tfrac{{\bf q}}{2} s\rangle \nonumber \\
&= \frac{s}{\sinh \mu - i \sin \nu} \times \begin{cases} \cosh \mu \sin \nu , & s s' = + 1 , \\[1ex] i \sinh \mu \cos \nu , & s s' = - 1 , \end{cases} \\
&\langle {\bf k} + \tfrac{{\bf q}}{2} s'| k_x |  {\bf k} - \tfrac{{\bf q}}{2} s\rangle \nonumber \\
&= \frac{q}{2} \frac{\sinh \mu \sin \nu}{\sinh \mu - i \sin \nu} \times \begin{cases} \sinh \mu , & s s' = + 1 , \\[1ex] - i \sin \nu , & s s' = - 1 , \end{cases} \\
&\langle {\bf k} + \tfrac{{\bf q}}{2} s'| k_y |  {\bf k} - \tfrac{{\bf q}}{2} s\rangle \nonumber \\
&= \frac{q}{2} \frac{\cosh \mu \cos \nu}{\sinh \mu - i \sin \nu} \times \begin{cases} \sinh \mu , & s s' = + 1 , \\[1ex] - i \sin \nu , & s s' = - 1 . \end{cases}
\end{align}
The square of the joint denominator in all of these expressions cancels with the Jacobian. Furthermore, note the distance to the focal points
\begin{align}
|{\bf k} + \frac{\bf q}{2}| &= \frac{q}{2} (\cosh \mu + \cos \nu) , \\
|{\bf k} - \frac{\bf q}{2}| &= \frac{q}{2} (\cosh \mu - \cos \nu) ,
\end{align}
which implies that the denominator in Eq.~\eqref{eq:intrinsic} only depends on $\mu$, and the $\nu$-integration in Eq.~\eqref{eq:intrinsic} can be performed directly. The subsequent $\mu$-integration is elementary but requires a cutoff $\Lambda = \hat{\Lambda}/\ell_B$ in momentum space to regulate the expression. The result of this calculation is
\begin{align}
\hat{K}_-^{00}(i \nu, {\bf x}) &= \frac{\pi x^2}{8 \sqrt{x^2 +  \nu^2}} , \\
\hat{K}_-^{xx}(i \nu, {\bf x}) &= \frac{\hat{\Lambda}}{2} - \frac{\pi \sqrt{x^2 +  \nu^2}}{8} , \\
\hat{\Delta}_-^{11}(i \nu, {\bf x}) &= \frac{3 x^2 \hat{\Lambda}}{32} - \frac{3 \pi x^2 \sqrt{x^2 +  \nu^2}}{128} , \\
\hat{\Delta}_-^{22}(i \nu, {\bf x}) &= \frac{x^2 \hat{\Lambda}}{32} - \frac{\pi x^2 \nu^2}{128 \sqrt{x^2 +  \nu^2}} .
\end{align}
Note that there is no intrinsic contribution to the mixed current-dipole response function ${\cal R}$.

\subsection{Extrinsic response}

Shifting the integration variables in Eq.~\eqref{eq:Kp}, the full extrinsic response reads
\begin{align}
&\hat{K}_{+}^{\mu\nu}(i \nu, {\bf x}) = \int \frac{d^2y}{(2\pi)^2} f(E_{+}({\bf y})) \nonumber \\*
&\quad \times\biggl[
\frac{{\hat F}_{\mu\nu}^{++}(-{\bf y}',-{\bf y})}{E_{+}({\bf y}) - E_{+}({\bf y}') - i \nu} 
- \frac{{\hat F}_{\mu\nu}^{++}({\bf y},{\bf y}')}{E_{+}({\bf y}') - E_{+}({\bf y}) - i \nu} \nonumber \\
&\quad
+ \frac{{\hat F}_{\mu\nu}^{-+}(-{\bf y}',-{\bf y}) }{E_{+}({\bf y}) - E_{-}({\bf y}') - i \nu}
- \frac{{\hat F}_{\mu\nu}^{+-}({\bf y},{\bf y}')}{E_{-}({\bf y}') - E_{+}({\bf y}) - i \nu}\biggr] . \label{eq:extrinsic}
\end{align}
Introducing polar coordinates for the $y$-integration, the response function is expressed as
\begin{align}
\hat{K}_{+}^{\mu\nu}(i \nu, {\bf x}) &= \int_0^1 dy \, y \, \hat{J}^{\mu\nu}_E(i \nu, y, x) ,
\end{align}
where $\hat{J}^{\mu\nu}_E(i \nu, y, x)$ is the angle integral of Eq.~\eqref{eq:extrinsic}. It is evaluated by transforming to a complex integration contour $z=e^{i\theta}$ around the unit circle. The numerator reads
\begin{align}
z^2 [(y\mp i\nu)^2 - |{\bf y} + {\bf x}|^2] &= - x y z (z-z_1) (z-z_2) ,
\end{align}
where the positions of the three simple poles are
\begin{align}
z_0 &= 0 , \\*
z_1 &= - \frac{x^2 + \nu^2 \pm 2 i y \nu}{2 x y} \nonumber \\*
&\qquad + \frac{1}{2 x y} \sqrt{(x^2+\nu^2) (x^2 - (2y\mp i\nu)^2)} , \\
z_2 &= - \frac{x^2 + \nu^2 \pm 2 i y \nu}{2 x y} \nonumber \\*
&\qquad - \frac{1}{2 x y} \sqrt{(x^2+\nu^2) (x^2 - (2y\mp i\nu)^2)} .
\end{align}
We have $z_1 z_2 = 1$ with $|z_1|<1$ and $|z_2| > 1$. The integral is evaluated by applying the residue theorem and picking up the two poles at $z_0$ and $z_1$ inside the contour. Performing the integration yields
\begin{figure}[t!]
\scalebox{0.8}{\includegraphics{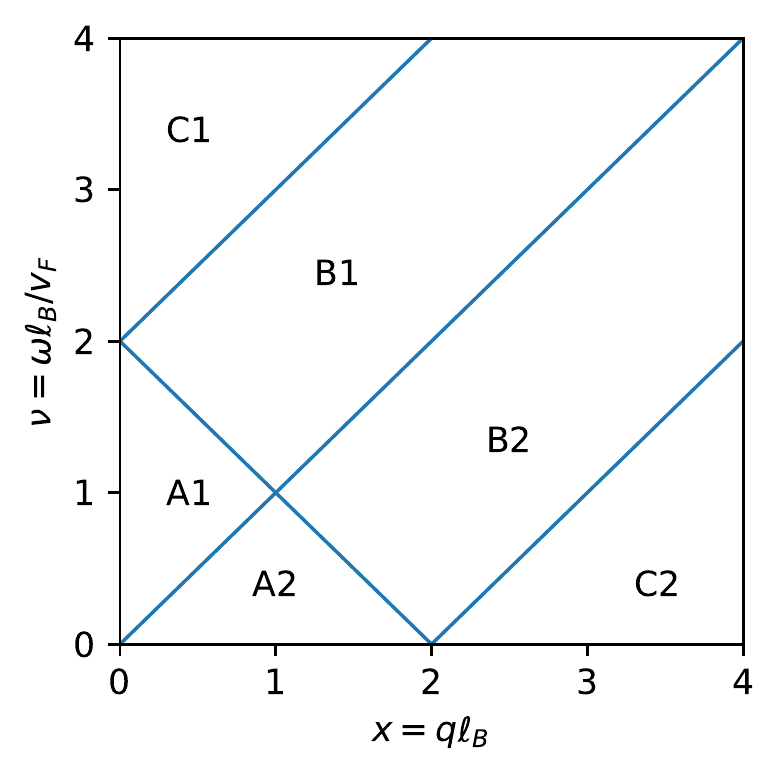}}
\caption{Frequency-momentum regions for the noninteracting Dirac response function.}
\label{fig:4}
\end{figure}
\begin{align}
&\hat{K}_{+}^{00}(i \nu, x)= - \frac{1}{2}  \nonumber \\*
&+ \frac{x^2}{8 \sqrt{x^2+\nu^2}} \biggl[ z \sqrt{1-z^2} + \arcsin z \biggr]_{-i\nu/x}^{(2-i\nu)/x} + (\nu \to - \nu) , \\
&\hat{K}_{+}^{xx}(i \nu, x) = - \frac{\nu^2}{2 x^2} \nonumber \\*
&+ \frac{\sqrt{x^2+\nu^2}}{8} \biggl[ z \sqrt{1-z^2} - \arcsin z \biggr]_{-i\nu/x}^{(2-i\nu)/x} + (\nu \to - \nu) , \\
&\hat{\Delta}_{+}^{11}(i \nu, x) = \frac{x^2}{8 \sqrt{x^2+\nu^2}} \nonumber \\*
&\biggl[ \frac{1}{4} z (2 z^2 - 1) \sqrt{1-z^2} + \frac{1}{4} \arcsin z \biggr]_{-i\nu/x}^{(2-i\nu)/x} + (\nu \to - \nu) , \\
&\hat{\Delta}_{+}^{22}(i \nu, x) \nonumber \\*
& = \frac{x^2}{8 \sqrt{x^2+\nu^2}} \biggl[ \frac{2}{3} (z^2 - 1) \sqrt{1-z^2}\biggr]_{-i\nu/x}^{(2-i\nu)/x} + (\nu \to - \nu) , \\
&\hat{\cal R}_{+}^{02}(i \nu, x) = \frac{x^2}{8 \sqrt{x^2+\nu^2}} \nonumber \\*
&\biggl[ \frac{1}{4} z (2z^2-5) \sqrt{1-z^2} - \frac{3}{4} \arcsin z \biggr]_{-i\nu/x}^{(2-i\nu)/x} + (\nu \to - \nu) , \\
& \hat{\cal R}_{+}^{x1}(i \nu, x) = (x^2+\nu^2) \, \hat{\cal R}_{+}^{02}(i \nu, x) .
\end{align}

\begin{figure*}
\scalebox{0.5}{\includegraphics{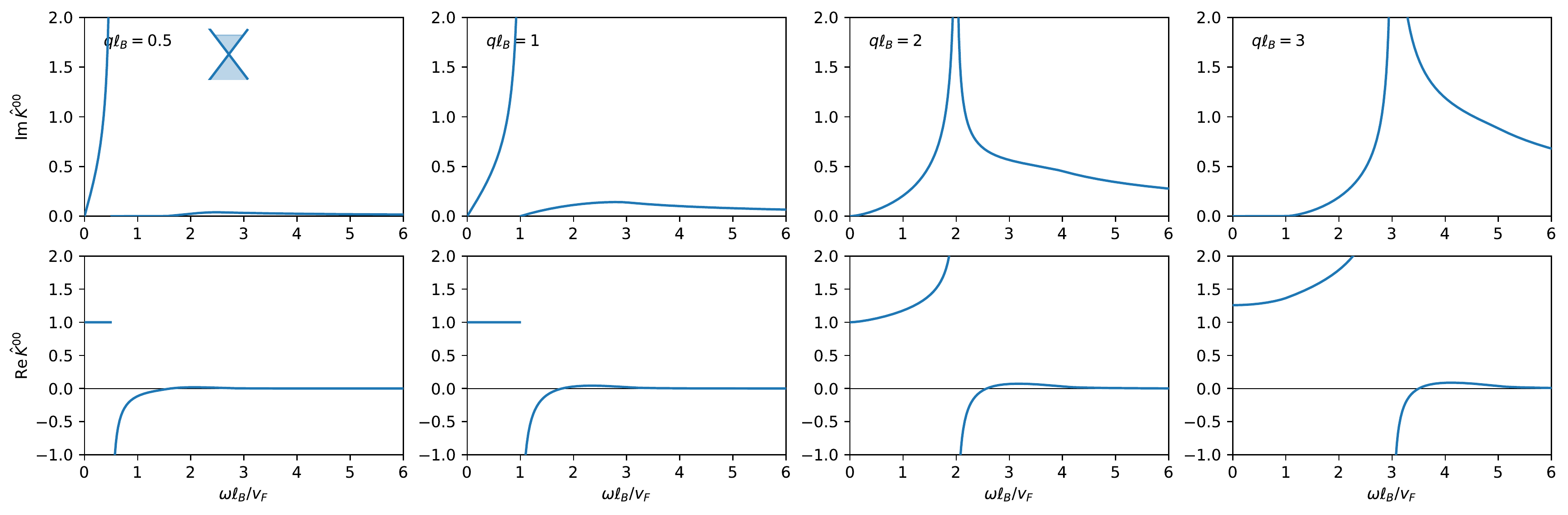}}
\caption{Dimensionless density response function of Dirac electrons for four different momenta (left to right panel) $q\ell_B = 0.5,1,2$, and $3$. The top panels show the imaginary part and the bottom panels the real part.}
\label{fig:5-1}
\end{figure*}

\subsection{Results}

In this section, we perform the analytic continuation to real frequencies $i\nu \to \nu + i0$ of the results for the intrinsic and extrinsic response functions computed in the previous two section. We parametrize our results as shown in Fig.~\ref{fig:4}. We first state the full result for the four response functions $K^{00}, K^{xx}, \Delta^{11}$, and $\Delta^{22}$:

\begin{figure*}
\scalebox{0.5}{\includegraphics{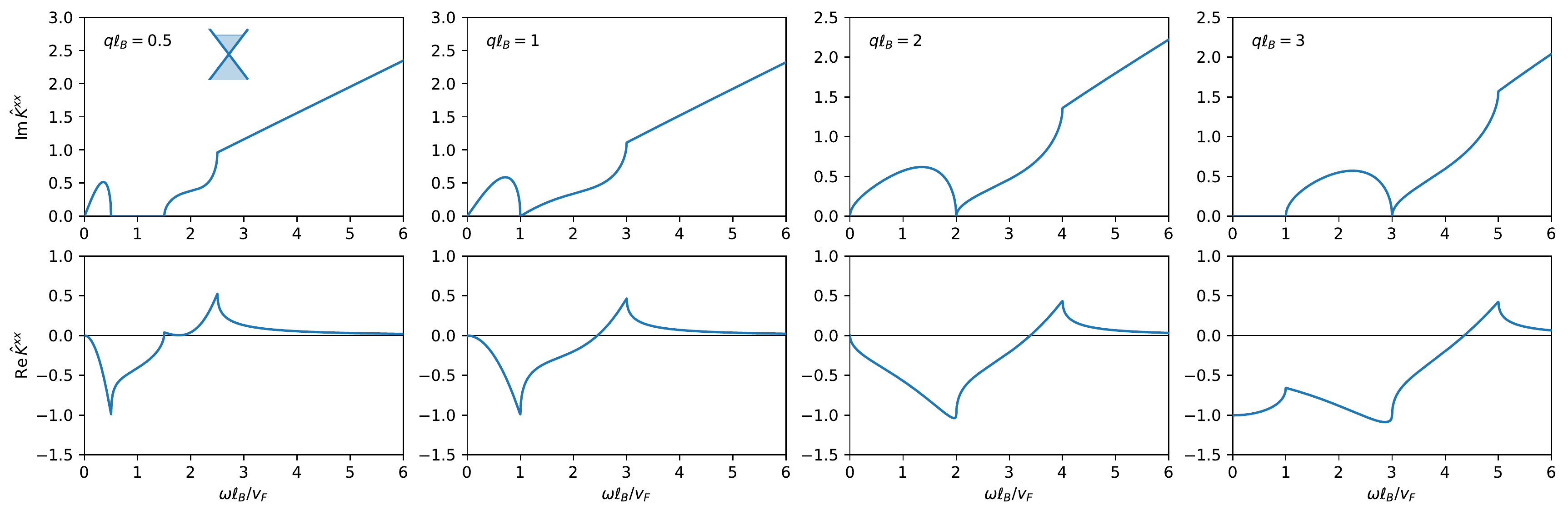}}
\caption{Dimensionless transverse current response function of Dirac electrons for four different momenta (left to right panel) $q\ell_B = 0.5,1,2$, and $3$. The top panels show the imaginary part and the bottom panels the real part.}
\label{fig:5-2}
\end{figure*}

\begin{align}
&{\rm Re} \hat{M}^{\alpha\alpha}(\nu, {\bf x})
\nonumber \\*
&= - h^\alpha(\nu,x) - f^\alpha(\nu,x)
\begin{cases}
g^\alpha_>\bigl(\frac{2+\nu}{x}\bigr) - g^\alpha_>\bigl(\frac{2-\nu}{x}\bigr) & {\rm A1} , \\[1ex]
g^\alpha_>\bigl(\frac{2+\nu}{x}\bigr) & {\rm B1} , \\[1ex]
g^\alpha_>\bigl(\frac{2+\nu}{x}\bigr) - g^\alpha_>\bigl(\frac{\nu-2}{x}\bigr) & {\rm C1} , \\[1ex]
\pi & {\rm A2} , \\[1ex]
\dfrac{\pi}{2} + g^\alpha_<\bigl(\frac{2-\nu}{x}\bigr) & {\rm B2} , \\[1ex]
g^\alpha_<\bigl(\frac{2+\nu}{x}\bigr) + g^\alpha_<\bigl(\frac{2-\nu}{x}\bigr) & {\rm C2} ,
\end{cases}
\end{align}
\begin{align}
&{\rm Im} \hat{M}^{\alpha\alpha}(\nu, {\bf x}) \nonumber \\*
&= - f^\alpha(\nu,x)
\begin{cases}
\pi & {\rm A1} , \\[1ex]
\dfrac{\pi}{2} + g^\alpha_<\bigl(\frac{2-\nu}{x}\bigr) & {\rm B1} , \\[1ex]
0 & {\rm C1} , \\[1ex]
- g^\alpha_>\bigl(\frac{2+\nu}{x}\bigr) + g^\alpha_>\bigl(\frac{2-\nu}{x}\bigr) & {\rm A2} , \\[1ex]
- g^\alpha_>\bigl(\frac{2+\nu}{x}\bigr) & {\rm B2} , \\[1ex]
0 & {\rm C2} ,
\end{cases} \label{eq:ImKdiag}
\end{align}
with
\begin{align}
h^\alpha(\nu,x) &= \begin{cases}
-1 & \alpha = 0 , \\[1ex]
\frac{\nu^2}{x^2} & \alpha = x , \\[1ex]
x^2 + \frac{\nu^2 (\nu^2 + 4)}{x^2} - 2 (\nu^2 + 1) & \alpha = 1 , \\[1ex]
-2 - \frac{4 \nu^2}{x^2} - \frac{\nu^4}{x^2} & \alpha = 2 ,
\end{cases} \\[2ex]
f^\alpha(\nu,x) &= \frac{1}{8 \sqrt{|\nu^2 - x^2|}} \begin{cases}
x^2 & \alpha = 0 , \\[1ex]
\nu^2 - x^2 & \alpha = x , \\[1ex]
\frac{3}{4} x^2 (\nu^2 - x^2) & \alpha = 1 , \\[1ex]
\frac{1}{4} x^2 \nu^2 & \alpha = 2 ,
\end{cases} \\[1ex]\nonumber
\end{align}
and
\begin{align}
g^\alpha_<(z) &= \begin{cases}
\arcsin z + z \sqrt{1-z^2} & \alpha = 0 , \\[1ex]
\arcsin z - z \sqrt{1-z^2} & \alpha = x , \\[1ex]
\arcsin z - \frac{1}{3} z (2z^2-5) \sqrt{1-z^2} & \alpha = 1 , \\[1ex]
\arcsin z + z (2 z^2 - 1) \sqrt{1-z^2} & \alpha = 2 ,
\end{cases} \\[1ex]
g^\alpha_>(z) &=  \begin{cases}
 - {\rm arccosh} \, z + z \sqrt{z^2-1} & \alpha = 0 , \\[1ex]
 - {\rm arccosh} \, z - z \sqrt{z^2-1} & \alpha = x , \\[1ex]
 - {\rm arccosh} \, z - \frac{1}{3} z (2z^2-5) \sqrt{z^2-1} & \alpha = 1 , \\[1ex]
 - {\rm arccosh} \, z + z (2 z^2 - 1) \sqrt{z^2-1} & \alpha = 2 .
\end{cases} 
\end{align}

\begin{figure*}
\scalebox{0.5}{\includegraphics{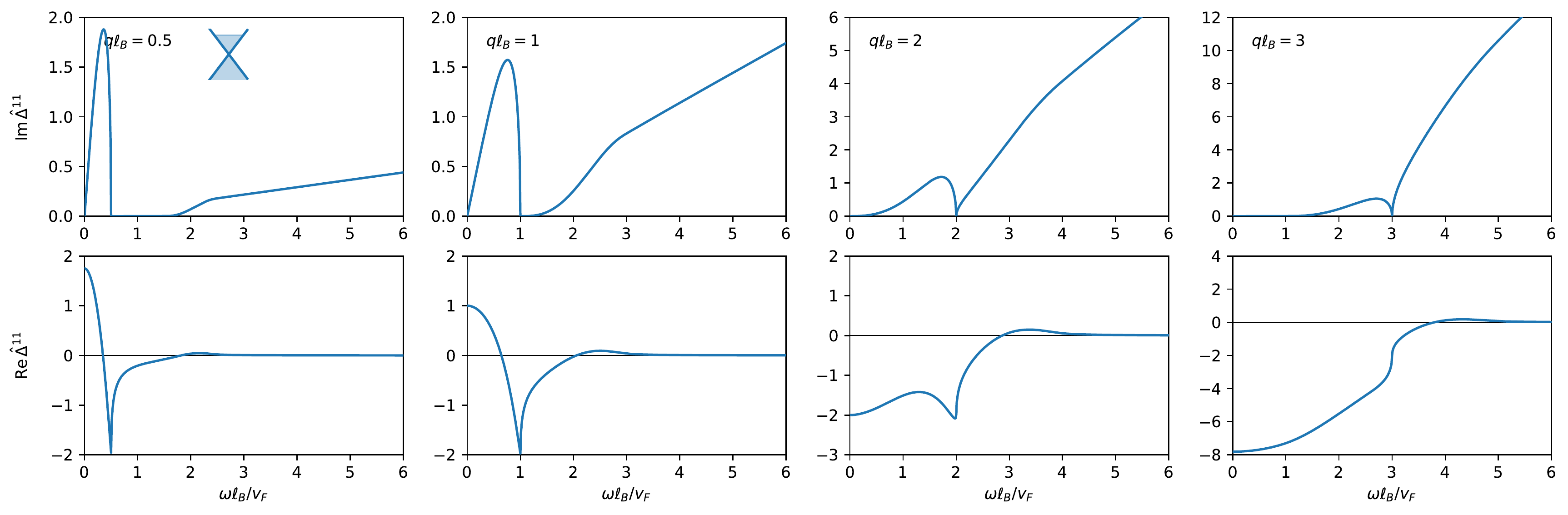}}
\caption{Dimensionless longitudinal dipole response function of Dirac electrons for four different momenta (left to right panel) $q\ell_B = 0.5,1,2$, and $3$. The top panels show the imaginary part and the bottom panels the real part.}
\label{fig:5-3}
\end{figure*}

\begin{figure*}
\scalebox{0.5}{\includegraphics{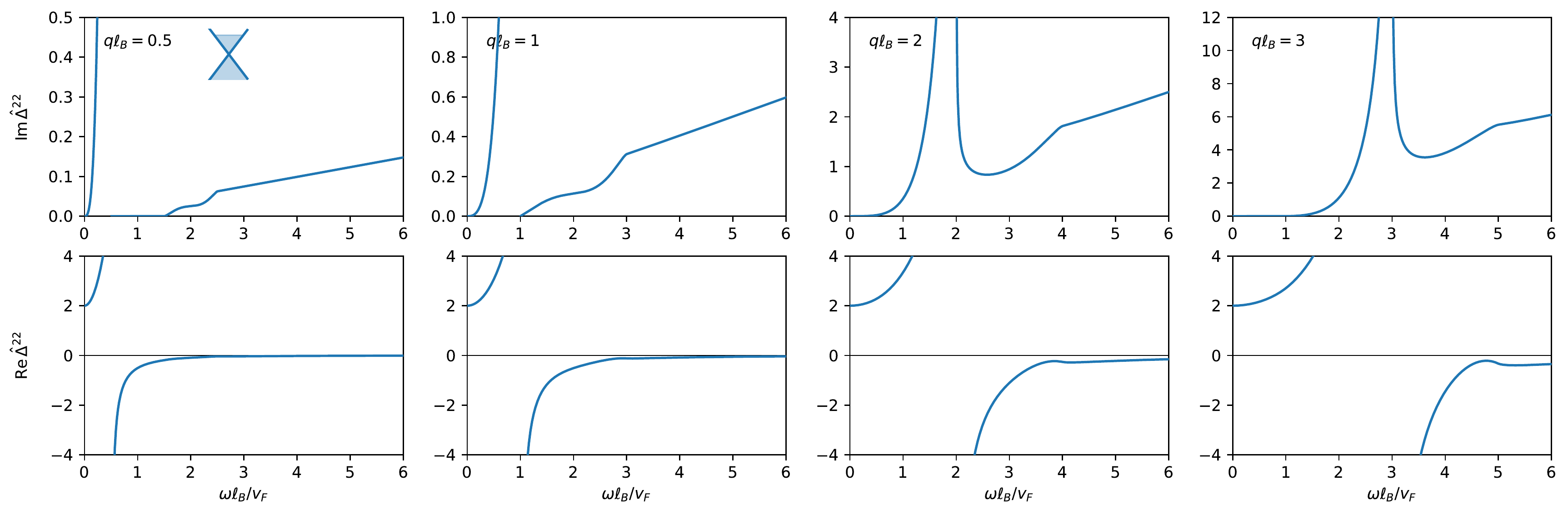}}
\caption{Dimensionless transverse dipole response function of Dirac electrons for four different momenta (left to right panel) $q\ell_B = 0.5,1,2$, and $3$. The top panels show the imaginary part and the bottom panels the real part.}
\label{fig:5-4}
\end{figure*}

\begin{figure*}
\scalebox{0.5}{\includegraphics{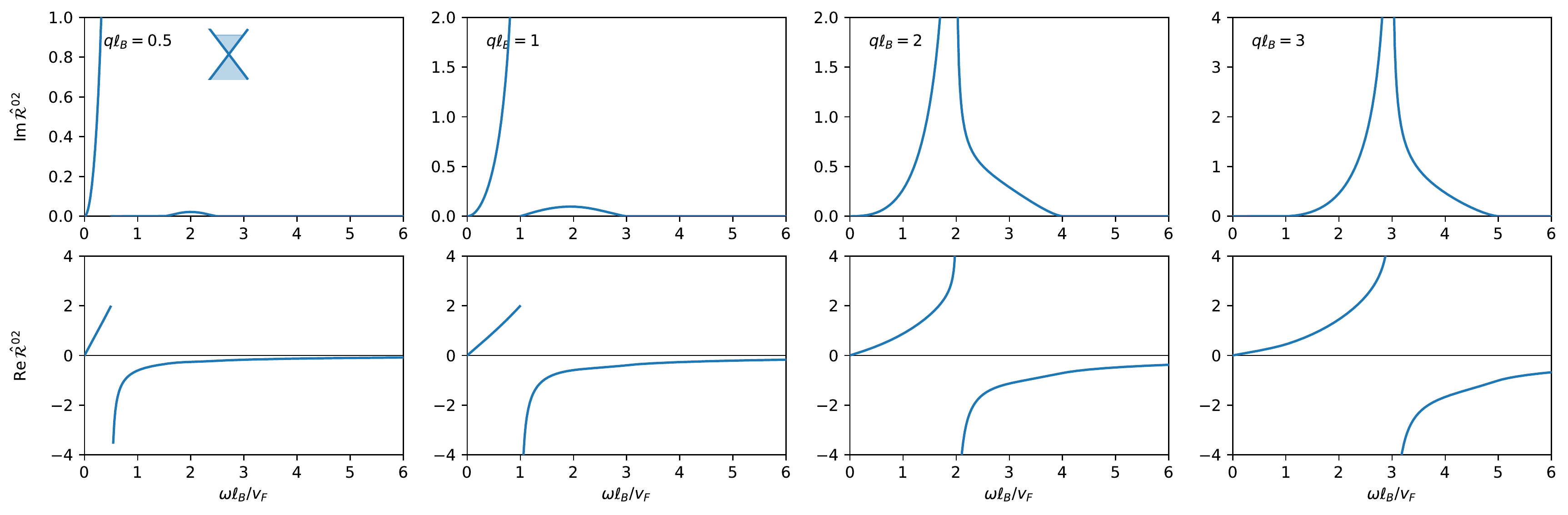}}
\caption{Dimensionless density-dipole response function of Dirac electrons for four different momenta (left to right panel) $q\ell_B = 0.5,1,2$, and $3$. The top panels show the imaginary part and the bottom panels the real part.}
\label{fig:5-5}
\end{figure*}

Taking the limit of  small momentum gives the results stated in Eqs.~\eqref{eq:K00q0},~\eqref{eq:Kxxq0},~\eqref{eq:K11q0}, and~\eqref{eq:K22q0} in the main text. 
Figures~\ref{fig:5-1}--\ref{fig:5-4} show the four response functions $K^{00}$, $K^{xx}$, $\Delta^{11}$, and $\Delta^{22}$ as a function of frequency for four momenta $q\ell_B = 0.5, 1, 2$, and $3$, where top panels show the imaginary part and bottom panels show the real part. 
The Dirac density response $K^{00}$ and transverse current response $K^{xx}$ function have been computed before in the graphene literature~\cite{hwang07,wunsch06,principi09}. Our results agree with these works (where we identify $k_F=1/\ell_B$ and $E_F = \hbar v_F/\ell_B$). 

\begin{figure*}[t]
\scalebox{0.5}{\includegraphics{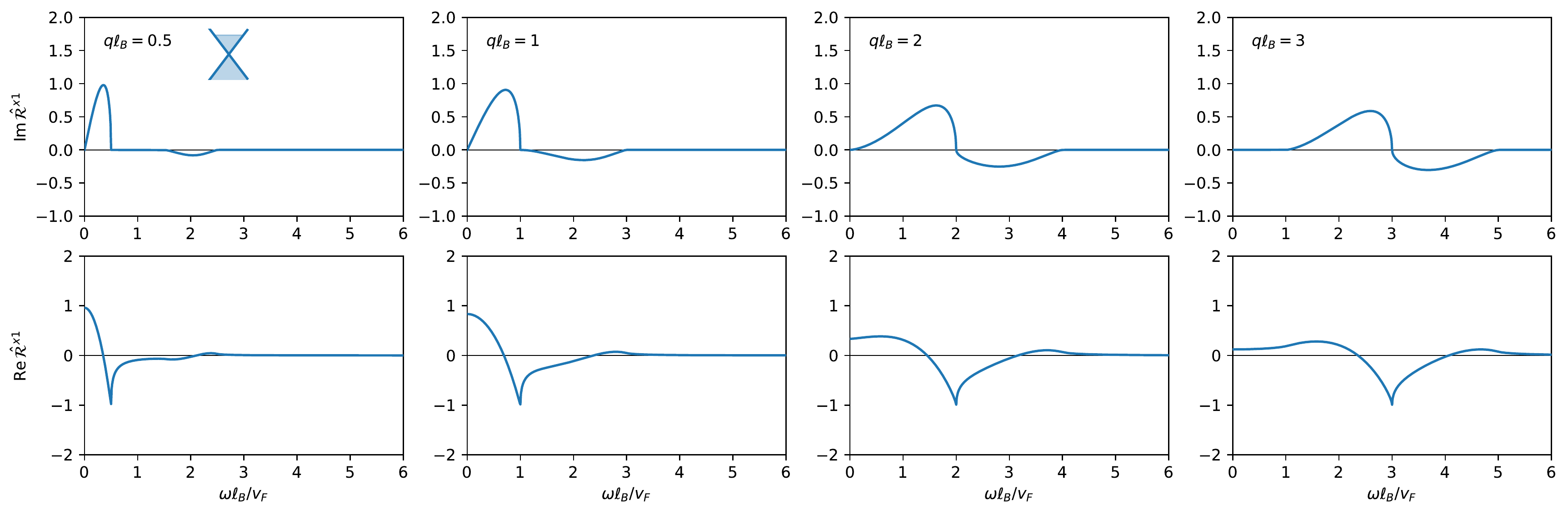}}
\caption{Dimensionless current-dipole response function of Dirac electrons for four different momenta (left to right panel) $q\ell_B = 0.5,1,2$, and $3$. The top panels show the imaginary part and the bottom panels the real part.}
\label{fig:5-6}
\end{figure*}

The result for the remaining two response functions ${\cal R}^{02}$ and ${\cal R}^{x1}$ is
\begin{align}
&{\rm Re} \hat{\cal R}^{\beta}(\nu, {\bf x}) \nonumber \\*
&= - h^\beta(\nu,x) - f^\beta(\nu,x)
\begin{cases}
g^\beta_>\bigl(\frac{2+\nu}{x}\bigr) - g^\beta_>\bigl(\frac{2-\nu}{x}\bigr) & {\rm A1} , \\[1ex]
g^\beta_>\bigl(\frac{2+\nu}{x}\bigr) & {\rm B1} , \\[1ex]
g^\beta_>\bigl(\frac{2+\nu}{x}\bigr) + g^\beta_>\bigl(\frac{\nu-2}{x}\bigr) & {\rm C1} , \\[1ex]
0 & {\rm A2} , \\[1ex]
g^\beta_<\bigl(\frac{2-\nu}{x}\bigr) & {\rm B2} , \\[1ex]
g^\beta_<\bigl(\frac{2+\nu}{x}\bigr) + g^\beta_<\bigl(\frac{2-\nu}{x}\bigr) & {\rm C2} ,
\end{cases}
\end{align}
\begin{align}
&{\rm Im} \hat{\cal R}^\beta(\nu, {\bf x}) \nonumber \\*
&= - f^\beta(\nu,x)
\begin{cases}
0 & {\rm A1} \\[2ex]
g^\beta_<\bigl(\frac{2-\nu}{x}\bigr) & {\rm B1} , \\[1ex]
0 & {\rm C1} , \\[1ex]
- g^\beta_>\bigl(\frac{2+\nu}{x}\bigr) + g^\beta_>\bigl(\frac{2-\nu}{x}\bigr) & {\rm A2} , \\[1ex]
- g^\beta_>\bigl(\frac{2+\nu}{x}\bigr) & {\rm B2} , \\[1ex]
0 & {\rm C2} ,
\end{cases}
\end{align}
with
\begin{align}
h^\beta(\nu,x) &= \begin{cases}
- \frac{2 \nu}{x} - \frac{\nu (\nu^2 - x^2)}{6 x} & \beta = 02 , \\[1ex]
- 1 + \frac{2 \nu^2}{x^2} + \frac{(x^2 - \nu^2)^2}{6 x^2} & \beta = x1 ,
\end{cases} \\
f^\beta(\nu,x) &= \dfrac{1}{8 \sqrt{|\nu^2 - x^2|}} \begin{cases}
x^2 \nu & \beta = 02 , \\[1ex]
x (x^2 - \nu^2) & \beta = x1 ,
\end{cases} \\
g^\beta_<(\nu,x) &= \frac{2}{3} (z^2 - 1) \sqrt{1-z^2} , \\
g^\beta_>(\nu,x) &= \frac{2}{3} (z^2 - 1) \sqrt{z^2-1} .
\end{align}
Figures~\ref{fig:5-5} and \ref{fig:5-6} show the response functions ${\cal R}^{02}$ and ${\cal R}^{x1}$ as a function of frequency for four momenta $q\ell_B = 0.5, 1, 2$, and $3$, where top panels show the imaginary part and bottom panels show the real part. Note the limiting value at small momentum 
\begin{align}
\hat{\cal R}^{02}(\nu, {\bf x}) &= - \dfrac{x}{\nu} - \frac{x^3}{\nu^3} \frac{\nu^2 - 3}{\nu^2 - 4} , \\
\hat{\cal R}^{x1}(\nu, {\bf x}) &= \frac{x^2}{\nu^2 (\nu^2-4)} ,
\end{align}
from which the scaling of the residue terms $Z_L, Z_T = {\cal O}(q^4)$ at long wavelengths is directly apparent.

\subsubsection{Small-argument scaling limit}\label{sec:scalinglimits}

In the small-frequency region and small momentum region (A1 and A2), introduce the scaling variable $\nu = s x$:
\begin{align}
\hat{K}^{00}(s x, {\bf x}) &= \begin{cases}
1 + i \dfrac{s}{\sqrt{1 - s^2}} & {\rm A1} , \\[2ex]
1 - \dfrac{s}{\sqrt{s^2-1}} & {\rm A2} ,
\end{cases} \\[1ex]
\hat{K}^{xx}(s x, {\bf x}) &= \begin{cases}
-s^2 + i s \sqrt{1-s^2}
 & {\rm A1} , \\[2ex]
-s^2 + s \sqrt{s^2-1} & {\rm A2} ,
\end{cases} \\[1ex]
\hat{\Delta}^{11}(s x, {\bf x}) &= \begin{cases}
2 - 4 s^2 + i 4 s \sqrt{1-s^2} & {\rm A1} , \\[1ex]
2 - 4 s^2 + 4 s \sqrt{s^2-1} & {\rm A2} ,
\end{cases} \\[1ex]
\hat{\Delta}^{22}(s x, {\bf x}) &= \begin{cases}
2+4 s^2 + i \frac{4 s^3}{\sqrt{1-s^2}} & {\rm A1} , \\[1ex]
2+4 s^2 - \frac{4 s^3}{\sqrt{s^2-1}} & {\rm A2} ,
\end{cases} \\[1ex]
\hat{\cal R}^{02}(s x, {\bf x}) &= \begin{cases}
2 s + i \frac{2 s^2}{\sqrt{1-s^2}} & {\rm A1} , \\[1ex]
2 s - \frac{2 s^2}{\sqrt{s^2-1}} & {\rm A2} ,
\end{cases} \\[1ex]
\hat{\cal R}^{x1}(s x, {\bf x}) &= \begin{cases}
1-2s^2 + i 2 s \sqrt{1-s^2} & {\rm A1} , \\[1ex]
1-2s^2 + 2 s \sqrt{s^2-1} & {\rm A2} .
\end{cases}
\end{align}
Taking the absolute value of the last two expressions gives the results for the residue terms~\eqref{eq:Zllimit} and~\eqref{eq:Ztlimit} stated in the main text. We also note the expansion of the inverse functions
\begin{align}
[\hat{K}^{xx}(s x, {\bf x})]^{-1} &= \begin{cases}
-1 - i \frac{\sqrt{1 - s^2}}{s} & {\rm A1} , \\[1ex]
-1 + \frac{\sqrt{s^2-1}}{s}  & {\rm A2} ,
\end{cases} \\
[\hat{K}^{00}(s x, {\bf x})]^{-1} &= \begin{cases}
1 - s^2 - i s \sqrt{1 - s^2} & {\rm A1} , \\[1ex]
1 - s^2 - s \sqrt{s^2 - 1} & {\rm A2} .
\end{cases}
\end{align}

\bibliography{bib_dirac}

\end{document}